%% file: main.tex
\documentclass[]{aastex631}

\usepackage{amsmath}
\usepackage{graphicx}
\usepackage{longtable}
\usepackage{booktabs}
\usepackage{multirow}
\usepackage{microtype}
\usepackage{soul}
\usepackage{algorithm}
\usepackage{algpseudocode}

\graphicspath{ {./figures} }

\begin{document}

\title{New metrics for identifying variables and transients in large astronomical surveys}

\correspondingauthor{Shih Ching Fu} \email{shihching.fu@postgrad.curtin.edu.au}

\author[0000-0002-0786-7307]{Shih Ching Fu}
\affiliation{International Centre for Radio Astronomy Research - Curtin University, GPO Box U1987, Perth, WA 6845, Australia}

\author[0000-0003-2506-6041]{Arash Bahramian}
\affiliation{International Centre for Radio Astronomy Research - Curtin University, GPO Box U1987, Perth, WA 6845, Australia}

\author[0000-0002-0637-7461]{Aloke Phatak}
\affiliation{ARC Centre for Transforming Maintenance through Data Science - Curtin University, GPO Box U1987, Perth, WA 6845, Australia}

\author[0000-0003-3124-2814]{James C. A. Miller-Jones}
\affiliation{International Centre for Radio Astronomy Research - Curtin University, GPO Box U1987, Perth, WA 6845, Australia}

\author[0000-0003-0052-128X]{Suman Rakshit}
\affiliation{School of Electrical Engineering, Computing and Mathematical Sciences, Curtin University, 
GPO Box U1987, Perth, WA 6845, Australia} 
\affiliation{Curtin Biometry and Agricultural Data Analytics, Centre for Crop and Disease Management, Curtin University, 
GPO Box U1987, Perth, WA 6845, Australia}

\author[0000-0003-2734-1895]{Alexander Andersson}
\affiliation{Astrophysics, Department of Physics, University of Oxford, Denys Wilkinson Building, Keble Road, Oxford OX1 3RH, UK}

\author[0000-0002-5654-2744]{Robert Fender}
\affiliation{Astrophysics, Department of Physics, University of Oxford, Denys Wilkinson Building, Keble Road, Oxford OX1 3RH, UK}

\author[0000-0002-6896-1655]{Patrick A. Woudt}
\affiliation{Department of Astronomy, University of Cape Town, Private Bag X3, Rondebosch 7701, South Africa}

\begin{abstract}
\input{abstract.tex}
\end{abstract}


\input{introduction.tex}
\input{methodology.tex}

\input{results.tex}
\input{discussion.tex}
\input{conclusion.tex}

\begin{acknowledgments}
This publication uses data generated via the \url{Zooniverse.org} platform, development of which is funded by generous support, including a Global Impact Award from Google, and by a grant from the Alfred P. Sloan Foundation. This research has made extensive use of arXiv and NASA's Astrophysics Data System Bibliographic Services.
\end{acknowledgments}

\vspace{5mm}
\facilities{MeerKAT}

\software{\textsc{numpy} \citep{2020NumPy-Array}, \textsc{scipy} \citep{2020SciPy-NMeth}, \textsc{matplotlib} \citep{hunter_matplotlib_2007}, \textsc{astropy} \citep{the_astropy_collaboration_astropy_2022}, \textsc{PyMC} \citep{wiecki_pymc-devspymc_2023}, \textsc{R} \citep{r_core_team_r_2023}, \textsc{Stan} \citep{stan_development_team_stan_2023}.}

\appendix
\input{appendix.tex}
\newpage
\bibliography{references}{}
\bibliographystyle{aasjournal}

\end{document}

%% file: abstract.tex
A key science goal of large sky surveys such as those conducted by the Vera C. Rubin Observatory and precursors to the Square Kilometre Array is the identification of variable and transient objects. One approach is analysing time series of the changing brightness of sources, namely, light curves. However, finding adequate statistical representations of light curves is challenging because of the sparsity of observations, irregular sampling, and nuisance factors inherent in astronomical data collection. The wide diversity of objects that a large-scale survey will observe also means that making parametric assumptions about the shape of light curves is problematic. We present a Gaussian process (GP) regression approach for characterising light curve variability that addresses these challenges. Our approach makes no assumptions about the shape of a light curve and, therefore, is general enough to detect a range of variable and transient source types. In particular, we propose using the joint distribution of GP amplitude hyperparameters to distinguish variable and transient candidates from nominally stable ones and apply this approach to 6394 radio light curves from the ThunderKAT survey. We compare our results with two variability metrics commonly used in radio astronomy, namely $\eta_\nu$ and $V_\nu$, and show that our approach has better discriminatory power and interpretability. Finally, we conduct a rudimentary search for transient sources in the ThunderKAT dataset to demonstrate how our approach might be used as an initial screening tool. Computational notebooks in \textsc{Python} and \textsc{R} are available to help deploy this framework to other surveys.

%% file: introduction.tex
\section{Introduction}
\label{sec:introduction}

Advancing our understanding of exotic astrophysical phenomena depends on accessing sufficiently large samples of such phenomena for analysis. As large survey projects such as the Rubin Observatory Legacy Survey of Space and Time (LSST;~\citealt{ivezic_lsst_2019}) and those planned for the Square Kilometre Array (SKA;~\citealt{dewdney_square_2009}) get commissioned and start monitoring vast portions of the sky, a key task is searching huge datasets for objects of interest. Indeed, surveys such as LSST and The HUNt for Dynamic and Explosive Radio transients with MeerKAT (ThunderKAT\footnote{\url{https://science.uct.ac.za/thunderkat}}; \citealt{fender_thunderkat_2017}) have science goals contingent on finding transient and variable phenomena. The investigation of high-energy transient phenomena, detected in x-ray, gamma-ray, radio, or other bands, helps us explore new physics under extreme conditions that cannot be reproduced in a laboratory. One discovery method is the statistical time series analysis of the changing brightness of sources, that is, their light curves. 

For identifying transient and variable sources, the statistical characterisation of their light curves must be descriptive and scalable: descriptive, in that we want to encode the properties of sources that best facilitate their identification and classification, and scalable, such that we can use these encodings across many surveys monitoring thousands to millions of sources~\citep{andersson_finding_2025}. This paper proposes a novel representation of light curves based on the hyperparameters derived by Gaussian process (GP) regression. We describe a new method for characterising sources with an emphasis on capturing the time-domain variability of a source, intending to identify transient and variable sources. 

There are many existing approaches for identifying and discovering specific subtypes of transient and variable objects. These typically take a template or matched filter approach, such as by~\cite{feng_matched_2017} in radio and~\cite{allen_findchirp_2012} in gravitational wave signals, which search for a specific shape or pattern of curve. Further complications exist, such as particular source types varying on particular timescales. For example, sub-second for fast radio bursts and pulsars, minutes for ultra-long period transients, hours to days for X-ray binaries and gamma-ray bursts, and weeks to months for tidal disruption events. However, in the context of commensal and serendipitous discovery in large surveys, no predetermined shape or time scale is available, so a general-purpose characterisation is needed. 

Our approach is therefore not tuned specifically to any particular source type but can fit models and quantify the variability of various light curve shapes. We envisage its use as an initial single-pass screening tool for rapidly reducing the search space of candidate sources without overly inflating the false negative rate. We demonstrate our approach on radio light curves and will apply it to survey data at other wavelengths in future work.

GPs have grown in popularity in the astrophysics literature, largely because they can be fitted to the sparse and unevenly sampled data typical of astronomical light curves. However, there are additional uses of GPs beyond resampling sparse data or fitting smooth curves. The parameters of GP regression models are physically interpretable, and their estimation easily fits into a Bayesian hierarchical framework such that observational uncertainties can be incorporated into every analysis stage. Therefore, we present our characterisation of light curves using GPs, striking a balance between oversimplified and overspecified models such that the results are statistically justified and astrophysically meaningful.

\subsection{Variability statistics in radio surveys}
\label{subsec:eta-V}

Numerous methods have been used to identify transient and variable sources amongst survey observations, but here we focus on those that calculate descriptive statistics for a light curve and proceed to flag sources that have extreme values of those statistics. For example, the Caltech-NRAO Stripe 82 Survey (CNSS; \citealt{mooley_caltech-nrao_2016}) identified variable sources by applying the Student-$t$ test to flux densities measured at different epochs with the null hypothesis that the source has not varied significantly between epochs. However, any selection bias in the choice of epochs for that comparison would have a strong effect on the success of this approach. The Murchison Widefield Array Transients Survey (MWATS;~\citealt{bell_murchison_2019}) used two metrics for identifying candidate variable objects: the time-averaged modulation index, $\overline{M}$, and the slope parameter of a linear model fitted to the time-averaged light curve, $\nabla_\textrm{S}$. A source was identified as a variable candidate if both metrics exceeded a threshold. However, since typical light curves are not adequately modelled as straight lines, model misspecification bias will cause an under-reporting of candidates.

Other large surveys conducted by radio telescopes such as the Australian Square Kilometre Array Pathfinder (ASKAP;~\citealt{murphy_askap_2021}), Murchison Widefield Array (MWA;~\citealt{bell_survey_2014}), Low-Frequency Array (LOFAR;~\citealt{swinbank_lofar_2015}), and MeerKAT~\citep{andersson_bursts_2023} have also used two measures: the flux density coefficient of variation, $V_\nu$, and the reduced $\chi^2$ statistic of variability, $\eta_\nu$, to quantify the significance of epoch-to-epoch variability of sources. The former is mathematically equivalent to the modulation index, $M$, and the latter is similar to a Student-$t$ test approach except that all data points are simultaneously compared with a reference value to produce a single statistic for the entire light curve. Since $V_\nu$ and $\eta_\nu$ are commonly used in radio astronomical surveys, we shall restate their formulations and use them to benchmark our approach.

For a light curve that consists of $N$ time-ordered flux density measurements $S_{i,\nu}$, $i = 1, \dots, N$, at frequency $\nu$ with uncertainties $\sigma_{i,\nu}$, $V_\nu$ is defined as the sample standard deviation $s$ divided by the overall mean $\overline{S_\nu}$:
\begin{equation} \label{eq:V_nu}
V_\nu = \frac{s}{\overline{S_\nu}} = \frac{1}{\overline{S_\nu}}\sqrt{\frac{N}{N-1}\left(\overline{S^2_\nu} - \overline{S_\nu}^2 \right)}
\end{equation}

The statistic $\eta_\nu$ is defined as:
\begin{equation} \label{eq:eta_nu}
\eta_\nu = \frac{1}{N-1} \sum_{i=1}^{N} \frac{\left(S_{i,\nu} - \overline{S_\nu}^*\right)^2}{\sigma_{i, \nu}^2} \sim \chi^2_{N-1}
\end{equation}
where $\overline{S_\nu}^*$ denotes the weighted mean such that for weights $w_{i, \nu} = 1/\sigma_{i, \nu}^2$:
\begin{equation}
\overline{S_\nu}^* = \frac{\sum_{i=1}^N w_{i, \nu} S_{i, \nu}}{\sum_{i=1}^N w_{i, \nu}}
\end{equation}

Following the guidance of \cite{bell_survey_2014}, sources with high values of both $V_\nu$ and $\eta_\nu$ are considered promising transient or variable candidates. A high value of $V_\nu$ suggests that the source was observed to brighten (or dim) to a large multiple (or fraction) of its usual level, and a high value of $\eta_\nu$ suggests that the observed size and number of variations in flux density are inconsistent with a source that is non-varying.

The main advantages of these two metrics are that they are easy to calculate and straightforward to deploy into the data processing pipeline of large surveys. However, as evident from Eqs.~\ref{eq:V_nu}--\ref{eq:eta_nu}, and in common with the methods mentioned earlier, none of these metrics incorporates any time information and, therefore, amount to summarising light curves into a coarse single-number statistic. This information loss dilutes their discriminatory power.

In contrast, some studies have used statistical learning for the automated clustering and classification of light curves \citep[e.g.,][]{rowlinson_identifying_2019}. Such methods have proven effective when confined to the specific survey from which the training data were obtained; however, they are susceptible to overfitting and cannot be easily translated to new surveys without being re-trained. 

To avoid both oversimplification and overspecification, we propose using GPs to characterise light curves. A GP representation of light curves will implicitly model the covariance between observations, the temporal information that a $(\eta_\nu, V_\nu)$ parameterisation discards. Furthermore, careful choice of covariance kernel functions will mitigate overfitting. Finally, light curve fits from a Bayesian hierarchical modelling (BHM) approach using GP priors will fully incorporate both observational and parameter estimation uncertainties into any posterior predictions. 

\subsection{GPs in time-domain astronomy}

Developed by~\cite{krige_statistical_1951} for geostatistics, the interested reader is directed to~\citet[Ch.~2]{rasmussen_gaussian_2006} and~\citet[Ch.~4]{matheron_matherons_2019} for a full mathematical treatment of GP regression. Recently, Gaussian processes have become popular for analysing astronomical time-series~\citep{aigrain_gaussian_2023} with applications in estimating the value of astrophysical parameters, modelling components of variability in a source, and detecting specific phenomena. Examples of the latter are the numerous studies searching for exoplanets and stellar companions that have used GPs to model the variability of stellar light curves for determining rotation periods and radial velocities \citep{angus_inferring_2018, bortle_gaussian_2021, czekala_disentangling_2017}. Similarly, GPs have been used to model and detect (quasi-)periodicity in active galactic nuclei (AGN; \citealt{covino_detecting_2022, covino_looking_2020, kovacevic_two-dimensional_2020}). In gravitational wave research, \cite{moore_improving_2016} and \cite{demilio_density_2021} used GP modelling to estimate parameters such as chirp mass. Another common application of GPs is in data preprocessing before further scientific analysis. This includes the disentangling of `nuisance' signals, that is, signal components attributed to factors not of immediate scientific interest, such as in pulsar timing studies~\citep{van_haasteren_new_2014, antoniadis_international_2022}, or correcting for systematic oscillations due to mechanical faults~\citep{pope_transiting_2016}. Concerning sparse data, interpolating light curves using GP regression has improved the accuracy of automated supernovae classifiers~\citep{villar_superraenn_2020, stevance_what_2023}. 

However, in each of these applications, the GP model has been tuned to a specific type of signal or waveform. These models are therefore biased towards detecting particular source types at the expense of others. In this work, we propose a general-purpose model that can characterise variable sources that exhibit various behaviours, such as slowly evolving, bursting, persistent, and ultra-long periods.

Another contrast with the studies mentioned above is that we use GP hyperparameters as direct descriptors of light curves, instead of using GPs as only a means to interpolate poorly sampled observations. Similar to the approach of \cite{mclaughlin_using_2024}, who used raw hyperparameter values to characterise and classify AGN flares, we focus on using estimated hyperparameter values as the primary descriptors of light curves.

%% file: methodology.tex
\section{Data and Methodology} 
\label{sec:methodology}

\subsection{ThunderKAT data}
\label{sec:data}
We used the radio light curves of 6394 sources from commensal survey data captured by ThunderKAT\footnote{\url{https://science.uct.ac.za/thunderkat}}~\citep{fender_thunderkat_2017}. Between 2018 and 2022, ThunderKAT was scheduled on MeerKAT to observe X-ray binaries, Cataclysmic Variables, short Gamma-Ray Bursts and Type Ia Supernovae. For each of the targeted sources, multiple epochs of data were collected on the hundreds of radio sources found in the 1 deg$^2$ fields of view centred on these targets. The data used in this study were made available through the citizen science project by \cite{andersson_bursts_2023} and are publicly accessible online~\citep{andersson_anderssonastrobfs-mkt-analysis_2024}.

\begin{figure}
    \centering
    \includegraphics[width=\textwidth]{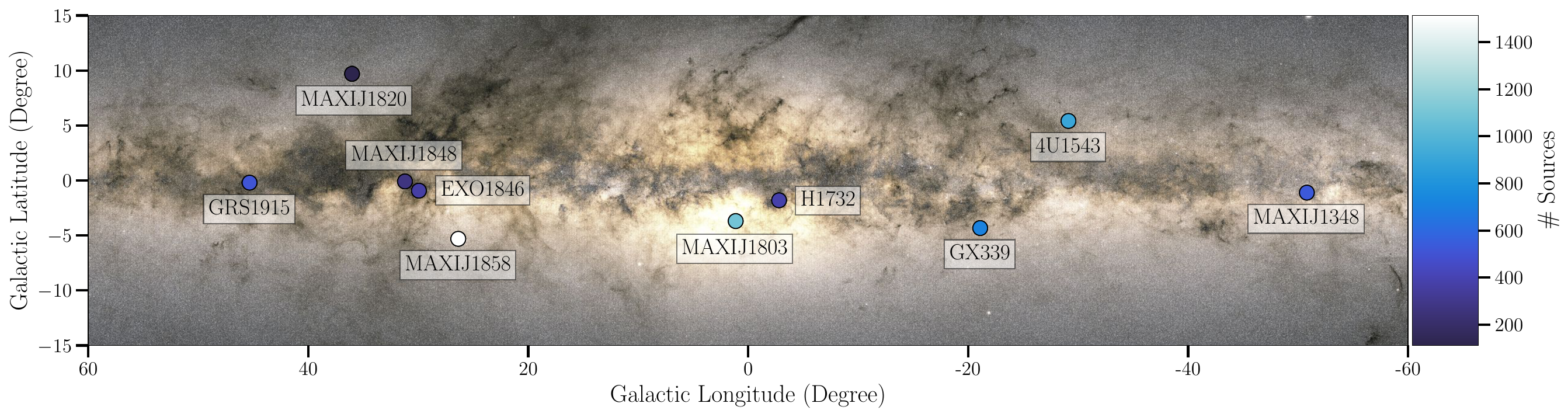}
    \caption{Projection onto the Galactic plane of the ThunderKAT fields of observation included in this work. The colour bar indicates the number of sources for which we have light curves in each field \citep{andersson_bursts_2023}. Fields are labelled according to their observation target. Background image from the Gaia mission (A. Moitinho; ESA/Gaia/DPAC).}
    \label{fig:skyplot}
\end{figure}

Data for each source consists of flux densities at 1.28 GHz (L-band) and their corresponding standard errors. These flux densities are peak values calculated within the synthesised beam of the source in radio images at each epoch, and the uncertainties were derived from fitting elliptical Gaussians to each island of pixels that constituted a source \citep{swinbank_lofar_2015}. Example light curves are shown in Figures~\ref{fig:postpreds_psds},~\ref{fig:egs-postpreds}, and~\ref{fig:candidates-postpred} and cutouts of typical radio images may be found in~\cite{rowlinson_identifying_2019},~\cite{andersson_serendipitous_2022} and~\cite{driessen_21_2022}. The number of points in the light curves ranged from 1--166 data points, and only detections are used in this analysis. Observations ranged from 1--1296 days. Furthermore, most of these light curves had multiple uneven gaps in their observations, lasting from 1--226 days. Recorded flux densities spanned from the noise limit of the telescope up to 1.24 Jy, with a similar range for the observational uncertainties\footnote{A small fraction of flux densities were $\lesssim 50~\mu$Jy -- below the approximate sensitivity limit of individual ThunderKAT observations -- and therefore likely to be spurious. Our analysis retained these since spurious measurements are typical of large surveys.}. As a commensal survey, all light curves are labelled with the name of the X-ray binary on which the field of view of the observation was centred (Figure~\ref{fig:skyplot}): 4U~1543$-$47 (\textsf{4U1543}), EXO~1846$-$031 (\textsf{EXO1846}), GRS~1915+105 (\textsf{GRS1915}), GX~339$-$4 (\textsf{GX339}), MAXI~J1848$-$015 (\textsf{J1848G}), Swift~J1858.6$-$0814 (\textsf{J1858}), MAXI~J1348$-$630 (\textsf{MAXIJ1348}), MAXI~J1803$-$298 (\textsf{MAXIJ1803}), H 1743-322 (\textsf{H1732}), and MAXI~J1820+070 (\textsf{MAXIJ1820}). 

We restricted our analysis to light curves with at least five data points, leaving $n = 5371$ light curves for analysis. This ensured sufficient information was available to fit a model and eliminated convergence problems in the fitting procedure. Before analysis, the flux density values were standardised by subtracting a light curve's sample mean and dividing it by its sample standard deviation. Similarly, the observational standard errors were standardised by dividing by the sample standard deviation. Consequently, all model fitting was performed in standardised units of source brightness before being transformed back into flux densities to report results. As described in Appendix~\ref{app:hyperpriors}, standardisation helps to improve the numerical stability for computation and simplifies the specification of prior distributions.

\subsection{Bayesian hierarchical model}

This section briefly describes the Bayesian hierarchical model (BHM) we have used to characterise light curves. We adopted the framework of \cite{wikle_2019}, which organises model hierarchies into three levels: a \emph{data model}, specifying the probabilistic model for the data; a \emph{process model}, which sets out the unobserved, or latent, process of interest; and a \emph{parameter model}, where the prior and hyperprior distributions are specified. These are described in Eqs.~\ref{eq:flux-distn} to~\ref{eq:ell-SE-M32-constraint}.

In Bayesian inference, we update prior distributions of the model parameters by using observed data to produce posterior distributions for each parameter. We then calculate the \emph{posterior predictive distribution} of fitted light curves from these posterior distributions. Curves sampled from this posterior predictive distribution may be compared to the originally observed light curve, providing a way to assess the plausibility of the fitted model. 

\subsubsection{Data model}
\label{subsec:data-model}

For each source in the ThunderKAT survey, we assumed its flux density $S_i$ at time $t_i$, $i = 1, \dots, N$, to be a Gaussian distributed random variable with a (latent) mean $f_i$ and variance $\hat{e_i}^2$ (Eq.~\ref{eq:flux-distn}). Note that the variance of each flux density measurement is not an estimated parameter but the reported standard error $\hat{e}_i$ from the observed light curve. 

\medskip
\noindent\fbox{\begin{minipage}{\textwidth}
\noindent \textbf{Data model}
\begin{equation} 
S_i \sim \mathcal{N}\left(f_i, \hat{e_i}^2\right), \quad i = 1, \dots, N. \label{eq:flux-distn}
\end{equation}

\noindent \textbf{Process model}
\begin{equation}
\boldsymbol{f} \sim \mathcal{GP}\left(\boldsymbol{0}, \boldsymbol{K}_{N \times N}\right) \label{eq:flux-gp}
\end{equation}
\begin{equation} 
    \begin{split}
[K]_{r,c} = \kappa(t_r,t_c \mid \boldsymbol{\theta}) =~& \sigma^2_\textrm{SE}\, \exp\left\{-\frac{1}{2}\frac{(t_r - t_c)^2}{\ell_\textrm{SE}^2}\right\} + \\
                                                                  & \sigma^2_\textrm{M32} \left( 1 + \sqrt{3}\,\frac{|t_r - t_c|}{\ell_\textrm{M32}}\right) \exp\left\{-\sqrt{3}\frac{|t_r - t_c|}{\ell_\textrm{M32}}\right\} + \\ 
                                                                  & \sigma^2_\textrm{P}\, \exp\left\{-\frac{2}{\ell_\textrm{P}^2} \sin^2\left( \pi\frac{|t_r - t_c|}{T}\right)\right\}, \qquad\qquad\qquad r,c = 1, \dots, N.  \label{eq:kappa}
\end{split}
\end{equation}

\noindent \textbf{Parameter model}
\begin{equation}
\sigma_\textrm{SE}, \sigma_\textrm{M32},\sigma_\textrm{P} \sim \mathcal{N}^+(0,1) \label{eq:sigma_prior}
\end{equation}
\begin{equation}
\ell_\textrm{SE}, \ell_\textrm{M32}, \ell_\textrm{P} \sim \mathrm{InvGamma}\left(\alpha = 3, \beta = \frac{1}{2} \times \lceil\textrm{range}(t)\rceil\right) \label{eq:ell_prior}
\end{equation}
\begin{equation}
T \sim \mathrm{Uniform}\left(2 \times \textrm{min}(\Delta t), \frac{1}{2} \times \textrm{range}(t)\right)  \label{eq:T_prior}
\end{equation}
\begin{equation}
\ell_\textrm{SE}, \ell_\textrm{M32},\ell_\textrm{P} > \textrm{min}(\Delta t) \label{eq:ell_min}
\end{equation}
\begin{equation}
\ell_\textrm{SE} > \ell_\textrm{M32}  \label{eq:ell-SE-M32-constraint}
\end{equation}

Bayesian hierarchical model used in this study. $S_i$ is the observed flux density of the $i$\textsuperscript{th} data point in a light curve whose true (but unobserved) flux density is $f_i$. The uncertainty in that data point is $\hat{e}_i$. $[K]_{r,c}$ is the covariance kernel of the Gaussian process prior for $f$, which is parameterised by $\boldsymbol{\theta}$ which is itself decomposed into three kernel terms: squared exponential (SE), Matern~3/2 (M32), and periodic (P). Each term has its own hyperparameters: $\sigma_\textrm{SE}$, $\sigma_\textrm{M32}$, and $\sigma_\textrm{P}$, are amplitude hyperparameters; $\ell_\textrm{SE}$, $\ell_\textrm{M32}$, and $\ell_\textrm{P}$, are length scale hyperparameters; and $T$ is the period hyperparameter. The quantity $\textrm{range}(t)$ is the total duration of a light curve, and $\textrm{min}(\Delta t)$ is the shortest time gap between points in a light curve.
\end{minipage}}

\subsubsection{Process model}
\label{subsec:process-model}
Since the latent $f$ is an unknown time-varying function, we model it as a stochastic process. More specifically, as shown in Eq. \ref{eq:flux-gp}, we model it as a Gaussian process (GP) defined with a zero mean function and an $N \times N$ covariance matrix $\boldsymbol{K}_{N \times N}$. Each $(r,c)$\textsuperscript{th} matrix element $[K]_{r,c}$ is specified by the kernel function $\kappa(t_r, t_c \mid \boldsymbol{\theta})$ in Eq. \ref{eq:kappa}. GPs are often used with a zero mean function, though prior knowledge of the underlying phenomena may justify specifying a parametric or non-parametric function instead \citep[p.~27]{rasmussen_gaussian_2006}. Depending on the choice of mean and kernel functions, a GP can generate flexible curves to fit various light curve characteristics such as high-frequency fluctuations, smooth monotonic trends, and quasi-periodicity. Our choices for the mean and kernel functions are justified below. 

\paragraph{Mean function} Surveys such as ThunderKAT contain thousands of sources whose light curves exhibit a wide range of behaviours~\citep{babu_skysurveys_2016}. These include long-term trends, broad peaks or troughs, sharp bursts, repeating patterns, and high-frequency oscillations. Examples of these are illustrated in Figures~\ref{fig:postpreds_psds},~\ref{fig:egs-postpreds} and~\ref{fig:candidates-postpred}. Furthermore, different fields of view have different numbers of pointings and data points. This means that no single parametric shape or pattern can be reliably assumed for an arbitrarily selected light curve. Consequently, in our GP regression model, we chose a zero mean function to avoid biasing our model toward any particular type of astrophysical phenomenon. Specifying a mean function of zero is consistent with having no specific knowledge about the shape or pattern of the modelled light curve\footnote{Standardised to have a mean of zero.} but does assume that the light curve has a stationary mean, that is, it neither trends upwards nor downwards. However, as shown in the top panels of Figure~\ref{fig:candidates-postpred}, some ThunderKAT light curves appear to brighten or dim during their observation. Nevertheless, some non-stationary behaviour can be addressed by using kernel functions tuned to appropriately long length scales. This study uses the squared exponential kernel term described below for this purpose.

In analyses of phenomena whose typical light curve shapes are known \emph{a priori}, a GP model can be employed to model the deviations from some underlying non-zero mean function. Examples include modelling the systematics in exoplanet studies such as that by \cite{pope_transiting_2016}, or the characterisation of quasi-periodic oscillations in X-ray light curves of gamma-ray bursts as done by \cite{song_detection_2024}. In this study, by specifying a zero mean function, we are directly modelling the observed flux densities and not the residuals.

\paragraph{Kernel function} The kernel function specifies the covariance relationship between data points in a Gaussian process. Properties of the kernel function, such as smoothness or periodicity, affect the final fit of the model, so the choice of kernel function strongly influences results \citep{stephenson_measuring_2022}. Following \cite{rasmussen_gaussian_2006}, we chose a composite kernel function comprising three distinct terms. Each constituent kernel term reflects a behaviour we expect to observe in light curves, such as smooth long-timescale trends, jagged short-timescale fluctuations, and, sometimes, periodicity. The covariance function $\kappa$ used in this analysis is described in Eq. \ref{eq:kappa} and consists of the sum of three kernel terms: squared exponential (SE), Matern~3/2 (M32), and periodic (P). These terms are a function of seven hyperparameters $\boldsymbol{\theta} = \left(\sigma_\textrm{SE}, \ell_\textrm{SE}, \sigma_\textrm{M32}, \ell_\textrm{M32}, \sigma_\textrm{P}, \ell_\textrm{P}, T\right)$. 

Trends and smooth variations at scales comparable to the total length of observation are modelled by the squared exponential kernel, whose length scale hyperparameter is constrained to favour that scale. This is a straightforward way to address non-stationarity in the mean, which cannot be accommodated by the zero mean function mentioned earlier. Jagged fluctuations are modelled by the Matern~3/2 kernel constrained to short length scales. Finally, periodic effects are encapsulated by the periodic kernel.

Including both a squared exponential kernel and a Matern~3/2 kernel term may introduce mathematical identifiability problems. This is partly due to the close functional relationship between these two kernels: the squared exponential kernel can be imagined as a Matern kernel of `infinite order', and the squared exponential kernel can be likened to a Matern kernel with extra smoothness constraints. Consequently, these kernels may recognise the same behaviours in a light curve, resulting in convergence problems or inflated uncertainty intervals. Details on how we mitigated this problem by imposing constraints on the hyperparameters are found in Appendix~\ref{app:hyperpriors}.

Lastly, we assumed that light curves could be treated as (second-order) stationary stochastic processes. That is, their covariance structures are only a function of the difference in time ($t - t'$) between data points. More intuitively, this means that regardless of which segment of a light curve is examined, its statistical properties are approximately the same as any other segment. However, verifying this assumption without knowledge of the underlying phenomenon proves challenging in astronomical contexts because data are often so sparse that metrics of non-stationarity are unreliable. The use of non-stationary GP kernels is a subject for future work.

\subsection{Parameter model}
\label{subsec:param-model}

Since a GP can be thought of as a prior distribution over functions, we will refer to the parameters of the kernel function as \emph{hyper}parameters and the prior distributions assigned to these as \emph{hyper}priors. These hyperpriors are shown in Eqs.~\ref{eq:sigma_prior}--\ref{eq:T_prior}. The notation $\textrm{range}(t)$ indicates the total duration of a light curve (in days), and $\textrm{min}(\Delta t)$ is the shortest time gap between points in a light curve (in days).

In addition to the choice of kernel function, the choice of hyperpriors will also influence the results of GP regression modelling. For example, if the data are sparse and the hyperpriors are strongly informative, these hyperpriors dominate the hyperparameter posterior distribution. Fortunately, it is straightforward to assess the influence of hyperpriors by comparing them with the posterior distribution of the hyperparameter, and a detailed justification of our choice of hyperpriors is provided in Appendix~\ref{app:hyperpriors}.

\subsection{Posterior distributions}
\label{subsec:posterior-dist}

The output of Bayesian inference is the joint distribution of all the parameters of interest. From this, we can derive estimates of particular hyperparameters expressed as posterior distributions. These distributions are useful and interpretable in their own right, but can also be used to generate so-called posterior predictive samples. These posterior predictive samples are curves that conform to the constraints of both the parameter model and the observed data and are useful for assessing the plausibility of model fit.

\subsubsection{Hyperparameter posteriors}

The posterior distributions of the seven hyperparameters in the GP regression model can be interpreted as descriptors of particular aspects of the original light curve. For example, if the posterior distribution of the length scale parameter for the squared exponential kernel $p(\ell_\textrm{SE}\mid y)$ has large values, it suggests that the light curve varies smoothly because of the strong autocorrelation between distant points. This can be extended to exploring the relationships between hyperparameter posteriors, which may reveal novel formulations for characterising astronomical sources\footnote{In this work, we have used the median of the hyperparameter posterior distributions.}. For example, in the bivariate case, GP regression modelling amounts to a dimension reduction procedure that situates every light curve in a 2-dimensional subspace of the 7-dimensional hyperparameter space where statistically (and perhaps astrophysically) similar sources might form clusters. In Section~\ref{subsec:var-heuristic}, we describe a two-dimensional subspace that partitions out variable sources and provides an alternative to using the $\eta_\nu$ and $V_\nu$ statistics discussed in Section~\ref{subsec:eta-V}.

\subsubsection{Posterior predictive curves}

After hyperparameter posterior distributions have been obtained for each fitted light curve, new (`predicted') curves $\boldsymbol{y^{*}}$ can be generated from the posterior predictive distribution $p(y^{*} \mid y)$. These predicted curves are conditioned on the observed data, and their distribution accounts for both the uncertainty in the data and in the estimated hyperparameters. \citet[p.~17]{rasmussen_gaussian_2006} derive an analytical expression for the posterior predictive distribution $p(y^{*} \mid y)$, which makes drawing these samples both straightforward and efficient. 

Posterior predictive curves are useful for assessing the fit of a GP regression model to the original data. This is done by inspecting the envelope of curves formed by repeated sampling of curves from the posterior predictive distribution, where the envelope's width reflects the uncertainty of the fit at any point along the curve. For example, in the left panels of Figure \ref{fig:postpreds_psds}, we observe that the 68\% quantile intervals of the posterior predictive samples encapsulate most of the original data (aside from the points with very large uncertainties), which suggests that the model is a plausible description of the data.

\subsubsection{Power spectral densities (PSD)}

In addition to providing a visual diagnostic for goodness-of-fit in the time domain, curves sampled from the posterior predictive distribution can be used to calculate power spectral densities (PSDs). Furthermore, we can estimate both the overall PSD and the individual contribution of each term in the kernel function. The relative contributions of the individual components can be compared by plotting them alongside the overall PSD, as in the right-hand panels of Figure~\ref{fig:postpreds_psds}. In these examples, the Matern~3/2 curve typically dominates the other two terms, and the very weak periodic kernel responses suggest an absence of periodicity.

Thus, inspecting the full and component-wise PSDs is another way to validate the model fit and scrutinise anomalies that might be difficult to identify in time-domain visualisations. This is particularly true for periodic patterns. However, care must be taken because the finite duration of light curves introduces fringing and other sampling artefacts into PSDs.

\subsection{Implementation}

The model fitting in this study was implemented in both \textsc{Python} and \textsc{R} to allow validation and facilitate wider deployment. It was implemented in \textsc{Python} (v3.10) with the \textsc{PyMC} library \citep[v5.6.1;][]{wiecki_pymc-devspymc_2023} using the default No U-turn Sampler \citep[NUTS;][]{hoffman_no-u-turn_2011}. It was implemented in \textsc{R} \citep[v4.4.1;][]{r_core_team_r_2023} and \textsc{CmdStan} \citep[v2.35.0;][]{stan_development_team_stan_2023} using the \textsc{cmdstanr} package \citep[v0.8.1;][]{gabry_cmdstanr_2024} and the default Hamiltonian Monte Carlo (HMC) sampler.

For both implementations, hyperparameter estimation was performed using chains of 2000 samples after discarding the first 2000  samples. Known as `burn-in' samples, these initial samples are discarded because they are likely biased due to the time it takes Monte Carlo (MC) chains to reach and explore the high-density regions of the parameter space. The rank-normalised $\hat{R}$ statistic \citep{Gelman_rhat_1992, vehtari_ranknormalization_2021} was used as a convergence diagnostic, and the effective sample size (ESS) was used to measure sampling efficiency. These metrics are provided by default in both \textsc{PyMC} and \textsc{CmdStan}. 

The computational cost of GP regression is known to scale at order $\mathcal{O}(n^3)$, where $n$ is the number of input data points. In practice, we observed that ThunderKAT light curves are sparse enough that it is not the regression computation that is the most time-consuming; rather, it is sampling the posterior predictive distribution that is the most onerous. This work has sped up this latter step by implementing a \emph{marginal} GP regression model as described in Appendix~\ref{app:marginalgp}.

%% file: results.tex
\section{Results} 
\label{sec:results}

In fitting the GP regression model described in Section~\ref{sec:methodology}, we segment the total variability exhibited by a light curve into components related to the amplitude and time scale of change. In the following sections, we present the results of light curve fitting, particularly the medians of posterior estimates of individual hyperparameters. All hyperparameters were estimated using Markov chain Monte Carlo (MCMC) sampling, and for 99\% of cases, we achieved reasonable convergence and efficiency measures of $\hat{R}\leq 1.003$ and bulk ESS $> 1753$, respectively. 

\subsection{Light curve fits}

Before using the hyperparameter posteriors, confirming the plausibility of the model fits is important. This is accomplished by visually inspecting plots of curves sampled from the posterior predictive distribution. Conveniently, the additive nature of our GP model means that we can also plot the PSD of any posterior predictive curve as the sum of its constituent kernel terms. Seeing the relative contributions of each term in the frequency domain complements the interpretation of the fitted model in the time domain.

Below we describe the results of fitting the GP regression model specified in Eqs.~\ref{eq:flux-distn}--\ref{eq:ell-SE-M32-constraint} to four ThunderKAT light curves. These were chosen because they differ by sampling rate, non-stationarity, presence of outliers, and distribution of observational errors. Each example highlights how our GP model accommodates a diversity of light curve features that may arise from astrophysical processes, data quality issues, or both. 

Table~\ref{tab:post_medians} lists the medians of the fitted posteriors for the seven fitted hyperparameters of each light curve, namely, $\{\sigma_\textrm{SE}, \ell_\textrm{SE}, \sigma_\textrm{M32}, \ell_\textrm{M32}, \sigma_\textrm{P}, \ell_\textrm{P}, T\}$. The table also lists their respective $\eta_{\nu}$ and $V_{\nu}$ statistics for comparison. As discussed in Appendix~\ref{app:hyperpriors}, amplitude posterior estimates with medians above $0.67$ warrant further interest and are highlighted in boldface in Table~\ref{tab:post_medians}. Further examples of model fits to known X-ray binary sources are included in Appendix~\ref{app:known-xrbs}.

\begin{figure}
    \centering
    \includegraphics[width=0.95\linewidth]{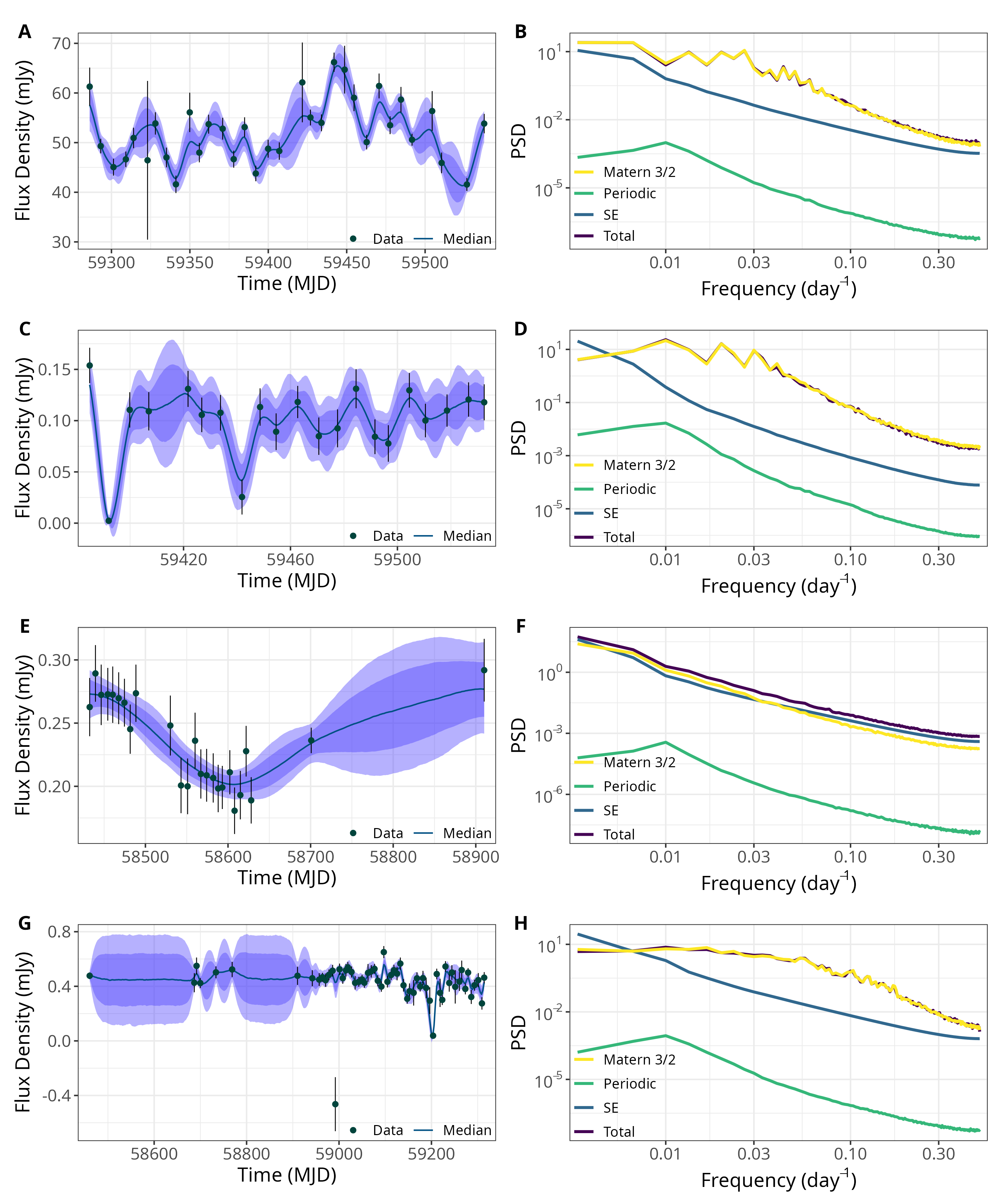}%
    \caption{Left: Posterior predictive samples of light curves in the fields around (\textsf{A}) \textsf{J1848G}, (\textsf{C}) \textsf{4U1543}, (\textsf{E}) \textsf{J1858}, and (\textsf{G}) \textsf{GRS1915}. Shaded regions are 68\% and 90\% quantile intervals, respectively. Standardised results have been transformed back to their original flux density scale. The inconsistent size of uncertainties is due to poor data quality in certain epochs. Right: Power spectral density (PSD) estimates of the posterior predictive curves in the left panel. Note that in \textsf{D} and \textsf{H}, numerical approximation artefacts misleadingly show the SE kernel exceeding the total PSD. See Table~\ref{tab:post_medians} for statistics of each source.}
    \label{fig:postpreds_psds} 
\end{figure}

The light curve in panel \textsf{A} of Figure~\ref{fig:postpreds_psds} features numerous undulations on the scale of $15-20$ days and is certainly a variable source. This is reflected in the estimated posterior median of the amplitude hyperparameter for the Matern-3/2 kernel, $\sigma_\textrm{M32} = 1.20$, which is greater than our nominal threshold of significance of $0.67$. Notice that the model has closely fitted these undulations and that most of the original observations lie within the 90\% uncertainty interval. Of note is the outlier near MJD 59324, which has a large observational uncertainty and is consequently ignored by the model in favour of its more precise neighbours. A similar relaxation of the model fit is observed around MJD 59424 and reflects how GP regression in a Bayesian framework directly incorporates uncertainties. Indeed, this is in common with other modelling approaches that weight the contribution of individual observations by their respective standard errors. These uncertainties are further propagated into the posterior estimates.

The PSD in the panel \textsf{B} of Figure~\ref{fig:postpreds_psds} confirms that most of the variability is explained by the Matern~3/2 kernel term. In contrast, the periodic component is very weak. This is consistent with the estimated amplitude posterior medians: $\sigma_\textrm{M32} = 1.20$, $\sigma_\textrm{SE} = 0.35$, $\sigma_\textrm{P} = 0.45$, where the former is above $0.67$ and the latter are below. This implies that the model has detected a non-trivial yet non-periodic correlation structure in the light curve. 

The light curve in Figure~\ref{fig:postpreds_psds}\textsf{C} is also regularly sampled but has fewer observations than the curve in panel \textsf{A}. This example shows that fitting to fewer observations results in wider uncertainty intervals. Again, there are two outliers, but these have smaller standard errors compared with the remaining data points than the outliers in panel \textsf{A}. Consequently, the model has deflected sharply towards these observations rather than ignoring them. Inspection of the radio image for that first apparent outlier revealed it as intrinsic variation in the source and not an image artefact to be ignored. The large median value of $\sigma_\textrm{M32} = 1.38$ again implies the presence of non-negligible variability, which contrasts with its small and unremarkable $\eta_\nu$ and $V_\nu$ statistics of $0.52$ and $0.27$, respectively.

Figure~\ref{fig:postpreds_psds}\textsf{E} shows a light curve that is unevenly sampled with widely spaced observations past MJD 58630. The model interpolates within these sparse data regions, but the uncertainty interval is wide. Overall, this light curve's first and second halves have a smooth down-then-upward trend. The moderately large estimates for amplitude ($\sigma_\textrm{SE} = 0.70$), and length-scale ($\ell_\textrm{SE} = 124.6$ days) hyperparameters encapsulate this behaviour. This is also reflected in the left-to-right downward slope in its PSD. Again, the globally calculated variability statistics of $(\eta_\nu = 0.94, V_\nu = 0.15)$ failed to detect the presence of this strong flux density inflection. 

The light curve in Figure~\ref{fig:postpreds_psds}\textsf{G} is of another unevenly sampled source with sparse regions where the uncertainty intervals are again very wide. Of particular interest are two outlying observations. The value observed just before MJD 59000 is clearly erroneous because flux densities must not be negative. The model mostly ignores this point because of its large standard error. In contrast, the other outlying point around MJD 59200, with its very small reported standard error, drags the model into a trough shape. The effect of this second outlier on inflating $\eta_\nu = 20.24$ and $V_\nu = 0.36$ is modest, making this light curve unremarkable according to these measures. The short time-scale, jagged behaviour exhibited in the latter half of this light curve is reflected in the hyperparameters of the Matern~3/2 kernel, namely a large $\sigma_\textrm{M32} = 1.21$ and short $\ell_\textrm{M32} = 17.3$ days. This behaviour is also captured in the PSD in panel \textsf{H}, where the Matern~3/2 kernel response dominates the squared exponential and periodic kernels.

\begin{table}
\caption{Fitted hyperparameter values for the example light curves shown in Figure~\ref{fig:postpreds_psds}. Medians of the hyperparameter posteriors are shown beside the variability statistics of $\eta_\nu$ and $V_\nu$. Boldface indicates median amplitudes greater than 0.67, a rule-of-thumb threshold above which a source is nominally labelled as a variable or transient.}
\label{tab:post_medians} 
\centering
\begin{tabular}{ccclccccccccc}
\toprule
\multirow{2}{*}{Figure~\ref{fig:postpreds_psds}}& \multirow{2}{*}{RA ($^\circ$)} & \multirow{2}{*}{Dec. ($^\circ$)} &\multirow{2}{*}{Field} & \multirow{2}{*}{$\eta_\nu$}& \multirow{2}{*}{$V_\nu$} & \multicolumn{7}{c}{Posterior Median} \\
& & & & & & $\sigma_\textrm{M32}$ & $\ell_\textrm{M32}$ & $\sigma_\textrm{SE}$ & $\ell_\textrm{SE}$ & $\sigma_\textrm{P}$ & $\ell_\textrm{P}$ & $T$ \\
\midrule
\textsf{A} & 281.9044 & -2.0325 & \textsf{J1848G} & 2.91 & 0.12 & \textbf{1.20} & 12.5 & 0.35 & 48.5 & 0.45 & 37.6 & 85.6 \\
\textsf{C} & 236.3573 & -48.1665 & \textsf{4U1543} & 0.52 & 0.27 & \textbf{1.38} & 7.74 & 0.36 & 28.8 & 0.47 & 22.0 & 50.1 \\
\textsf{E} & 284.4313 & -8.0519 & \textsf{J1858} & 0.94 & 0.15 & 0.56 & 73.4 & \textbf{0.82} & 117.3 & 0.53 & 72.39 & 165.0\\
\textsf{G} & 288.3110 & 11.4117& \textsf{GRS1915}  & 20.24	& 0.36	& \textbf{1.21}	& 17.3	& 0.31	& 171.7 & 0.42 & 128.6 & 292.2 \\
\bottomrule
\end{tabular}
\end{table}

\subsection{Hyperparameter posteriors}

\subsubsection{Squared exponential kernel}

\begin{figure}
    \centering
    \includegraphics[width=\textwidth]{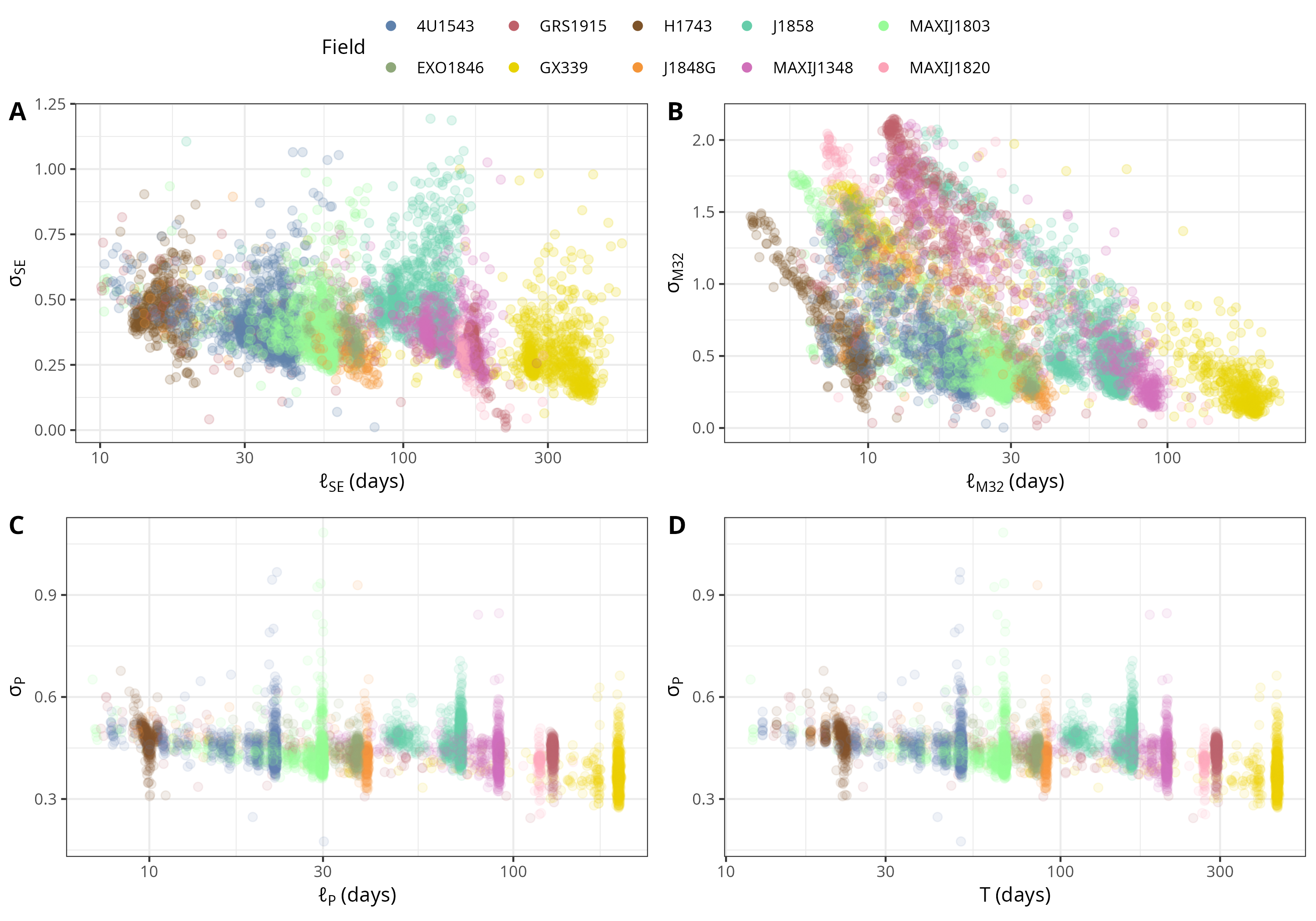}
    \caption{Posterior medians of our GP regression model's hyperparameters from each kernel term. \textsf{A}: Squared exponential (SE) kernel, \textsf{B}: Matern~3/2 (M32) kernel, \textsf{C} and \textsc{D}: Periodic (P) kernel. Each point corresponds to one light curve. Amplitudes, $\sigma$, are in standardised units, and length scales $\ell$ and period $T$ are in days on a logarithmic scale. Colours indicate the field in which the light curve was observed. Notice the distinct clustering associated with the field of observation.}
    \label{fig:hyperparam}
\end{figure}

Figure~\ref{fig:hyperparam}\textsf{A} shows the scatter plot of the two estimated hyperparameters of the squared exponential kernel, namely $\ell_\textrm{SE}$ and $\sigma_\textrm{SE}$. Recall that these are from the first term of the kernel function specified in Eq.~\ref{eq:kappa}. Each plotted point corresponds to the median of the posterior distribution for each hyperparameter of a single ThunderKAT light curve. The different colours indicate the X-ray binary in whose field the light curve was observed. An obvious pattern is the clustering by field, as evidenced by identically coloured points appearing close to each other. Sources observed in the same field tend to have similar hyperparameter posterior medians. This is particularly true for the length scale hyperparameter $\ell_\textrm{SE}$, where clusters usually span distinct ranges in that dimension. Some fields, such as \textsf{GX339} and \textsf{J1858}, also exhibit some internal clustering, which suggests that they, in turn, are composed of a mixture of subpopulations. 

We further observe that the clusters mentioned above are not as well separated in the vertical $\sigma_\textrm{SE}$ direction as in the horizontal $\ell_\textrm{SE}$ direction. Indeed, many different fields have overlapping ranges of amplitudes. In contrast, fields are much better delineated by length scale. For example, fields \textsf{4U1543} and \textsf{J1858} have clear separation in $\ell_\textrm{SE}$, but have ranges in $\sigma_\textrm{SE}$ that heavily overlap.

Visual inspection of the bivariate relationship between $\ell_\textrm{SE}$ and $\sigma_\textrm{SE}$ reveals no obvious correlation. Changes in $\ell_\textrm{SE}$ are not accompanied by a consistent change in $\sigma_\textrm{SE}$. However, with increasing $\ell_\textrm{SE}$, there does appear to be an increase in the variance of $\sigma_\textrm{SE}$ with a right skew --- amplitudes below 0.5 units are favoured across all fields and especially in \textsf{GX339}. However, within each field, the lack of strong patterns suggests that these two hyperparameters are mostly uncorrelated when conditioned by the field of observation.

One explanation for the clustering by field is the presence of confounding effects attributable to the field. For example, each field has its particular duration and cadence of observation. Factors such as observational cadence are nuisance contaminants in the characterisation of light curves since they are artefacts of the data collection process rather than properties of the source itself. These effects would bias any source identification results and should be removed. In Section~\ref{subsec:survey-effects}, we discuss our finding that two hyperparameters ($\sigma_\textrm{SE}$ and $\sigma_\textrm{M32}$) appear largely unaffected by field confounding.

\subsubsection{Matern~3/2 kernel}

The distribution of the hyperparameters for the Matern~3/2 kernel has some distinct differences from that of the squared exponential kernel. In Figure~\ref{fig:hyperparam}\textsf{B} there is a strong negative correlation between $\ell_\textrm{M32}$ and $\sigma_\textrm{M32}$. Increases in the length scale hyperparameter usually accompany decreases in the amplitude hyperparameter. This suggests that concerning the variability captured by the Matern~3/2 kernel term, there is a continuum of light curves ranging from smoothly undulating (small $\sigma_\textrm{M32}$, long $\ell_\textrm{M32}$) to steep and jagged (large $\sigma_\textrm{M32}$, short $\ell_\textrm{M32}$). Light curves at the latter extreme may constitute good candidates for transient sources.

As with the squared exponential hyperparameters, there is some clustering by field for the Matern~3/2 kernel hyperparameters, with similar proportions of overlap. This is again more associated with the length scale rather than the amplitude, which is better mixed. Indeed, even if the colouring in Figure \ref{fig:hyperparam} were removed, distinguishing different fields horizontally by clusters would still be possible visually. This suggests that, like the squared exponential kernel, the Matern~3/2 kernel term is still biased by field-specific effects.

\subsubsection{Periodic kernel}

\begin{figure}
    \centering
    \includegraphics[width=\textwidth]{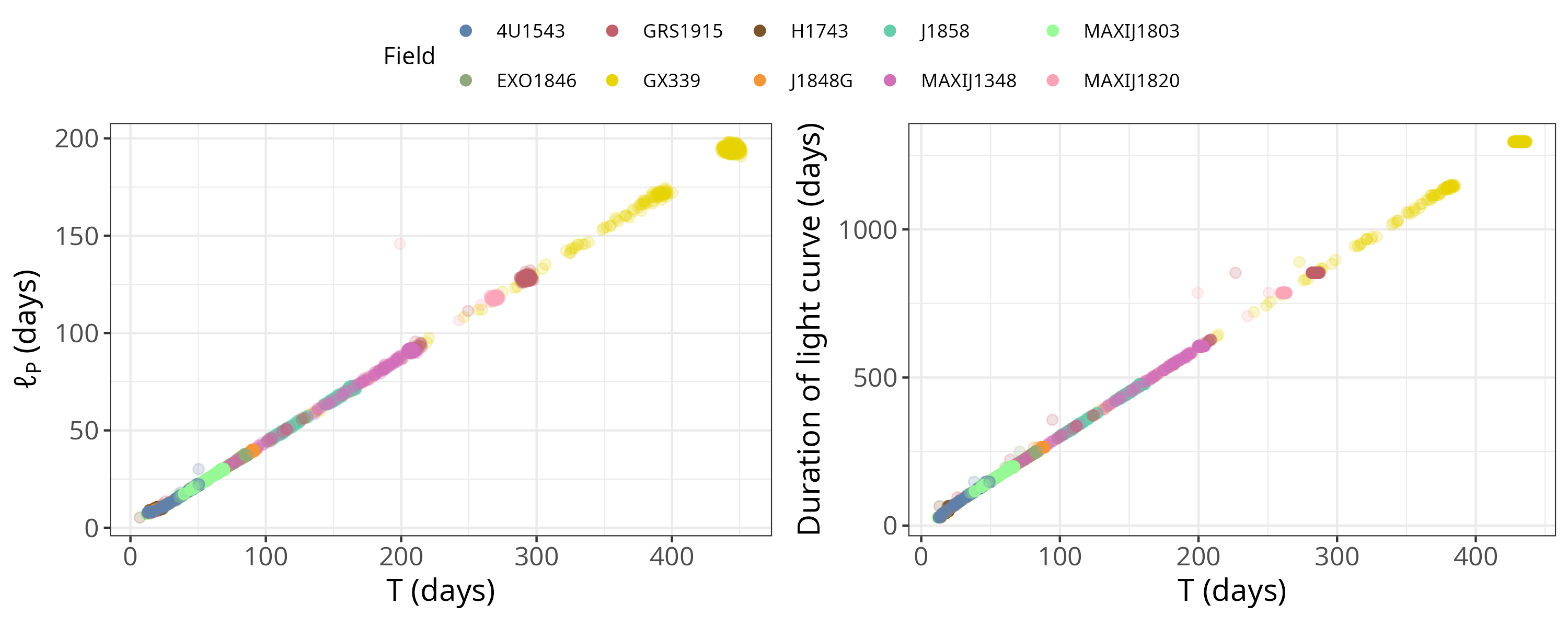}%
    \caption{Left: Scatter plot of the posterior medians of the period ($T$) and length scale ($\ell_\textrm{P}$) hyperparameters of the periodic kernel. These two quantities have a strong positive correlation ($R^2 = 0.996$). Right: Scatter plot of the posterior median of $T$ against the total duration of each light curve.}
    \label{fig:per-T-ell-T-N}
\end{figure}

Visual inspection of the two bivariate scatter plots between the posterior medians of the three hyperparameters of the periodic kernel (Figures~\ref{fig:hyperparam}\textsf{C} and~\textsf{D}) does not reveal any associations between the length scale and amplitude, or the period and amplitude hyperparameters. However, the similarity between these two plots suggests a strong correlation between length scale ($\ell_\textrm{P}$) and period ($T$). This is confirmed in the left panel of Figure~\ref{fig:per-T-ell-T-N} by the almost perfect positive correlation ($R^2 = 0.996$) between these two hyperparameters.

One explanation for this strong correlation is again a confounding factor. The right panel of Figure~\ref{fig:per-T-ell-T-N} is a plot of $T$ against the total duration of each light curve, the latter being a property of the data, not of the underlying source. This duration information has leaked into the hyperparameter posteriors through their respective hyperpriors (Eqs.~\ref{eq:ell_prior} and~\ref{eq:T_prior}) and fully accounts for the estimated values of $T$ and $\ell_\textrm{P}$. This implies no periodicity was detected; otherwise, shorter periods or length scales would have been found. 

%% file: discussion.tex
\section{Discussion} 
\label{sec:discussion}

\subsection{Amplitude hyperparameters as indicators of transience and variability}
\label{subsec:amplitude-hyperparams}

The novel aspect of this work is that we use the posterior medians of GP regression \emph{hyperparameters} as the primary descriptors of light curves instead of employing a GP model as an intermediary step in a pre-processing pipeline. In this section, we argue that of the seven hyperparameters estimated for each light curve, the amplitude hyperparameters of the Matern~3/2 and squared exponential kernels show the most promise as robust indicators of variability. The location of sources in the two-dimensional subspace of $(\sigma_\textrm{M32}, \sigma_\textrm{SE})$ is a useful cue for identifying variable and transient candidates from amongst other sources whose observations do not exhibit much evidence of change over time. In Appendix~\ref{app:PCA}, we justify why using only two of the seven hyperparameters is sufficient to discriminate between variable and non-variable sources.

\begin{figure}
\centering
\includegraphics[width=0.7\textwidth]{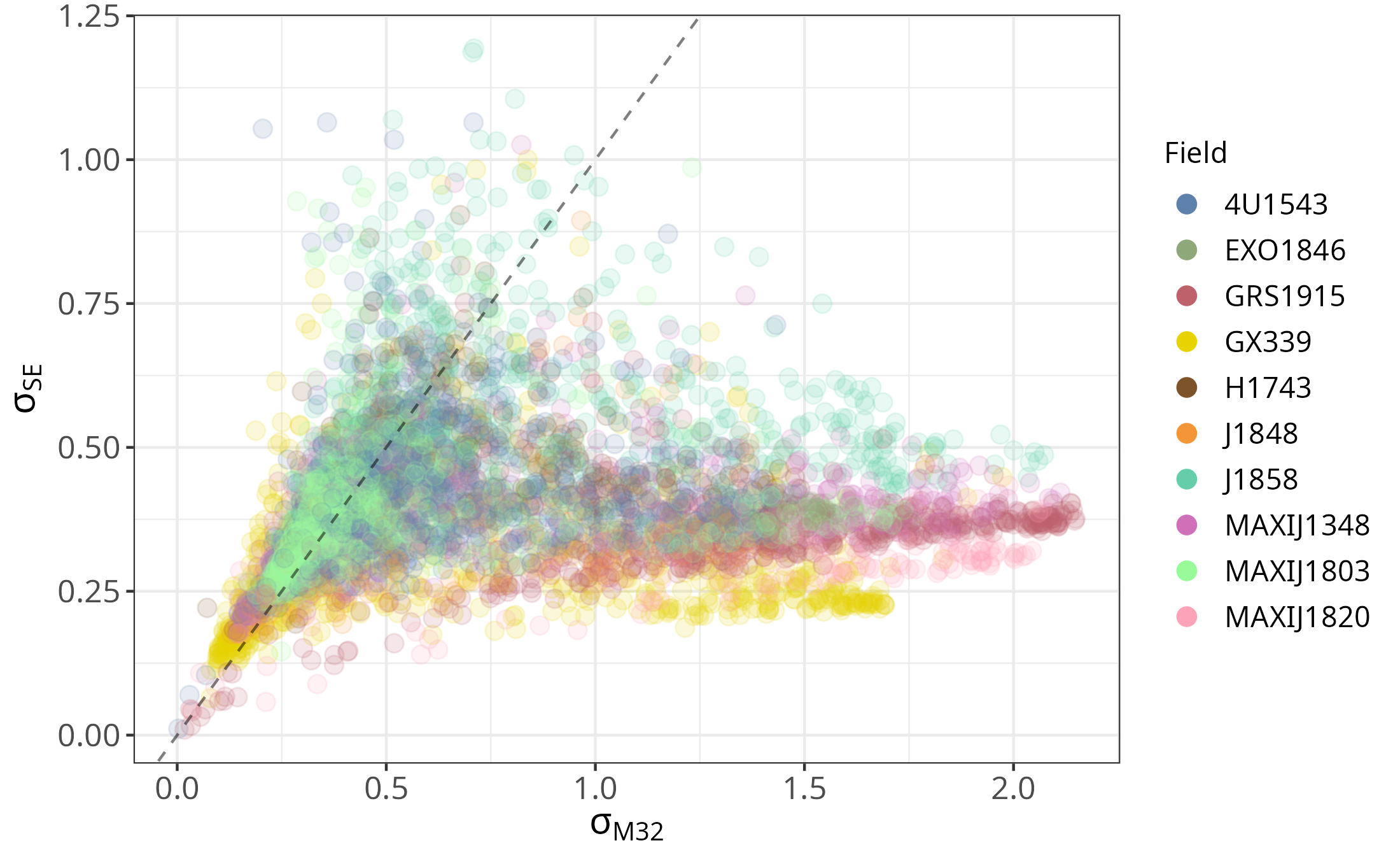}
\caption{\label{fig:sigma-M32-SE}Scatter plot of the posterior medians of the amplitude hyperparameters of the squared exponential ($\sigma_\textrm{SE}$) and Matern~3/2 ($\sigma_\textrm{M32}$) kernels for each fitted light curve. Amplitudes are in standardised units, and colours indicate the field where the light curve was observed. The dashed line indicates the line of equality between the two hyperparameters.}
\end{figure}

\subsubsection{Region of stable candidate sources}
\label{subsec:stable-sources}

\begin{figure}
\centering
\includegraphics[width=\textwidth]{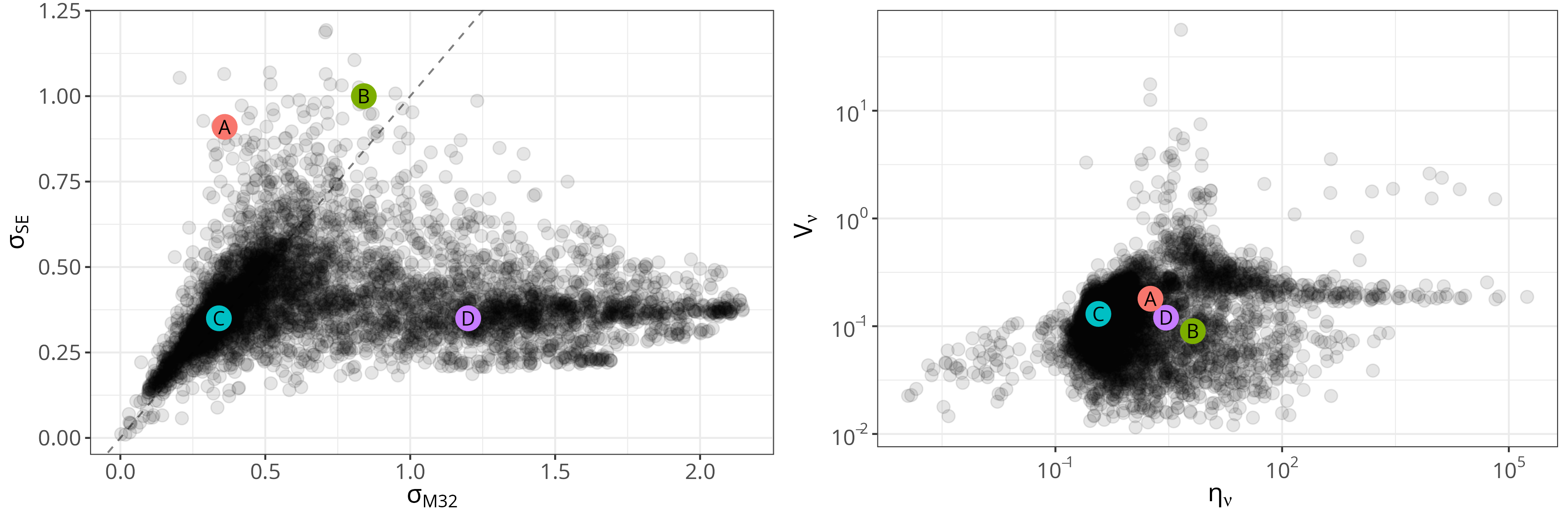}
\caption{\label{fig:sigma-M32-SE-examples}Left: Scatter plot of the posterior medians of the amplitude hyperparameter of the squared exponential and Matern~3/2 kernels, where each point corresponds to a ThunderKAT light curve. Amplitudes are in standardised units, and the dashed line indicates equality between these two hyperparameters. Right: Scatter plot of the $(\eta_\nu, V_\nu)$ statistics for each ThunderKAT light curve. The labelled points indicate where in the parameter space the four light curves shown in Figure~\ref{fig:egs-postpreds} reside. We observe that the four labelled light curves are better separated in $(\sigma_\textrm{M32}, \sigma_\textrm{SE})$-space than in $(\eta_\nu, V_\nu)$-space. See Table~\ref{tab:etanu-comparison} for metrics.}
\end{figure}

The scatter plot of $\sigma_\textrm{M32}$ against $\sigma_\textrm{SE}$ in Figure~\ref{fig:sigma-M32-SE} reveals a fan-shaped distribution with conspicuous clustering about the line of equality between these two hyperparameters. A representative light curve chosen from this elongated cluster, marked by the \textsf{C} in Figure~\ref{fig:sigma-M32-SE-examples}, is plotted in panel \textsf{C} of Figure~\ref{fig:egs-postpreds}. It shows a very flat fitted curve. Indeed, the inspection of light curves located in this region of the hyperparameter space reveals that they are all similarly non-varying or have observational uncertainties that are too large to classify them as variable or transient. Therefore, we have designated this region of the hyperparameter space as containing nominally `stable' sources.

Using the rule-of-thumb value of 0.67 as a threshold, the lower left quadrant of this hyperparameter space where $\sigma_\textrm{SE}, \sigma_\textrm{M32} \le 0.67$ covers approximately 58\% of the total number of light curves analysed in this study. We conjecture that this is a conservative lower bound on the proportion of sources observed in ThunderKAT that are neither variable nor transient. Sources in this quadrant should be de-prioritised in any search for transients. The true boundary of this region requires a more precise definition and could be approximated by bootstrapping the joint distribution of $(\sigma_\textrm{M32}, \sigma_\textrm{SE})$ from light curves of sources that are known to be non-varying.

\subsubsection{$\sigma_\textrm{M32}$ and $\sigma_\textrm{SE}$ as variability heuristics}
\label{subsec:var-heuristic}

\begin{figure}
\centering
\includegraphics[width=\textwidth]{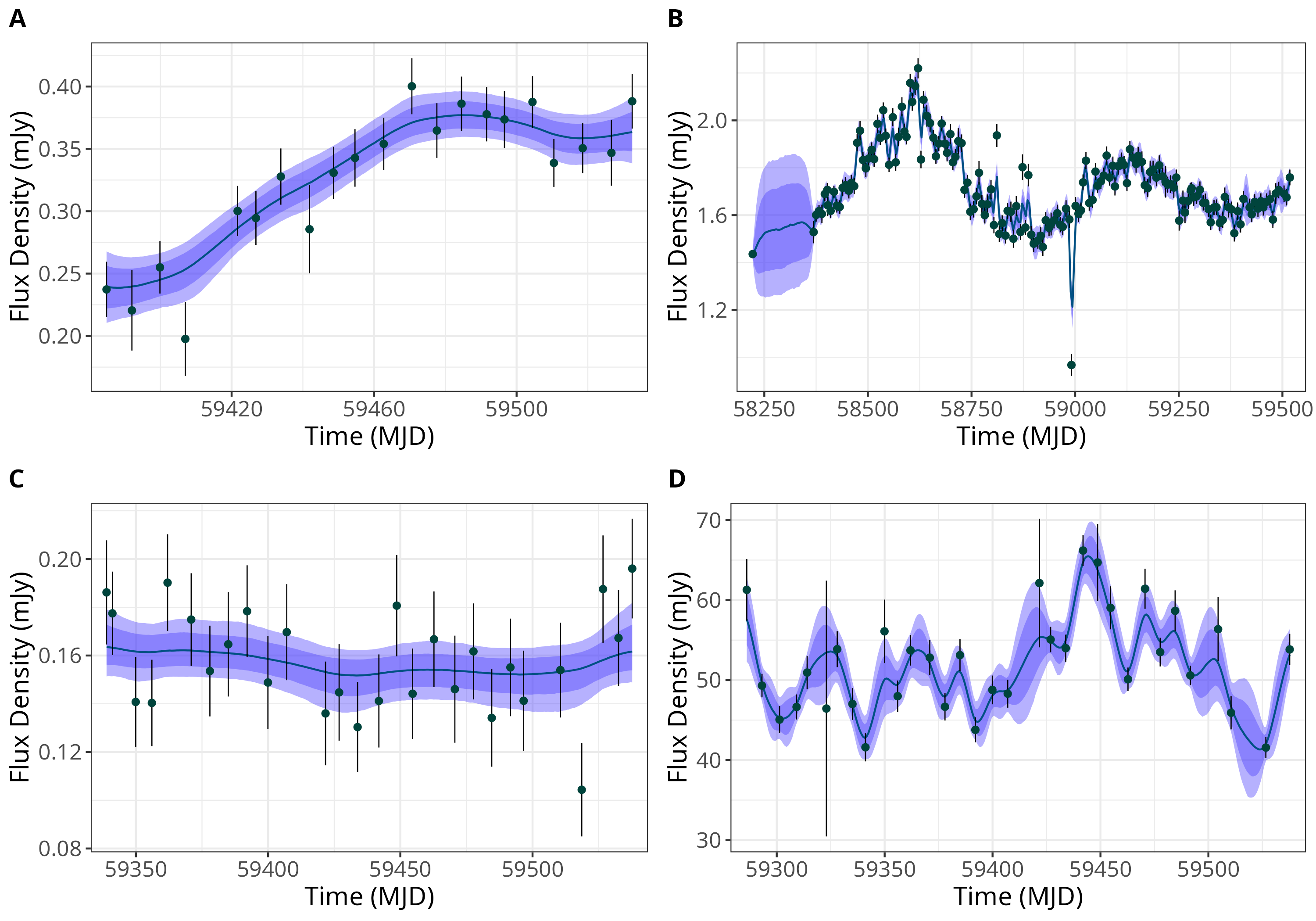}
\caption{\label{fig:egs-postpreds}Posterior predictive samples of four light curves from the ThunderKAT survey. The dark curve traces the median of 2000 posterior predictive samples, and the shaded regions form the 68\% and 90\% intervals. The points and vertical bars indicate the original observations and their respective uncertainties. Each light curve is drawn from a distinct region of the $(\sigma_\textrm{M32},\sigma_\textrm{SE})$ hyperparameter space as labelled in Figure~\ref{fig:sigma-M32-SE-examples}. \textsf{A}: $\sigma_\textrm{SE}$ is dominant. \textsf{B}: both hyperparameters are comparably large valued. \textsf{C}: both hyperparameters are small valued. \textsf{D}: $\sigma_\textrm{M32}$ is dominant. Specific coordinates and amplitude estimates are shown in Table~\ref{tab:etanu-comparison}.}
\end{figure}

If nominally stable sources populate the lower left quadrant of the $(\sigma_\textrm{M32}, \sigma_\textrm{SE})$ hyperparameter space, light curves of (statistically) varying sources should populate the other three quadrants. Visual inspection of light curves from those quadrants confirms this supposition, as shown in the remaining panels of Figure~\ref{fig:egs-postpreds}. In Figure~\ref{fig:egs-postpreds}\textsf{A}, the light curve shows a long timescale, smooth upward trend with few undulations, and is characterised by large $\sigma_\textrm{SE} = 0.91$ and small $\sigma_\textrm{M32} = 0.36$. In Figure~\ref{fig:egs-postpreds}\textsf{D}, this light curve with shorter time-scale variations has small $\sigma_\textrm{SE} = 0.35$ and large $\sigma_\textrm{M32} = 1.20$. Finally, in Figure~\ref{fig:egs-postpreds}\textsf{B}, where both hyperparameters are large, the light curve is a composite of very high frequency jittering superimposed onto intermediate timescale trends. These examples demonstrate that, in addition to screening out non-varying (or noisily indeterminate) sources, the magnitudes of the $\sigma_\textrm{SE}$ and $\sigma_\textrm{M32}$ amplitude hyperparameters are a useful heuristic for distinguishing between qualitatively different types of source behaviours.

\begin{table}
\caption{Amplitude hyperparameter values for the light curves shown in Figure~\ref{fig:egs-postpreds}. Medians of the amplitude hyperparameter posteriors are shown alongside the commonly reported variability statistics of $\eta_\nu$ and $V_\nu$. Boldface indicates median amplitudes greater than 0.67, a rule-of-thumb threshold above which a source might be considered variable or transient. All panels except for \textsf{C} would be potential transient or variable candidates.}
\label{tab:etanu-comparison}
\centering
\begin{tabular}{ccclcccc}
\toprule
Figure~\ref{fig:egs-postpreds} & RA ($^\circ$) & Dec. ($^\circ$) & Field & $\eta_\nu$ & $V_\nu$ & $\sigma_\textrm{M32}$ & $\sigma_\textrm{SE}$ \\
\midrule
\textsf{A} & 236.8389 & -47.2731 & \textsf{4U1543} & 1.81 & 0.18 & 0.36 & \textbf{0.91}\\
\textsf{B} & 255.1565 & -48.9462 & \textsf{GX339} & 6.59 & 0.09 & \textbf{0.84} & \textbf{1.00}\\
\textsf{C} & 270.3887 & -29.7998 & \textsf{MAXIJ1803} & 0.37 & 0.13 & 0.34 & 0.35\\
\textsf{D} & 281.9044 & -2.0325 & \textsf{J1848G} & 2.91 & 0.12 & \textbf{1.20} & 0.35\\
\bottomrule
\end{tabular}
\end{table}

\subsubsection{Comparison with labelled data}
\label{subsubsec:label_comparison}

To validate our claim that $(\sigma_\textrm{M32}, \sigma_\textrm{SE})$ are useful as variability heuristics, we compared our findings with source classifications made by citizen scientists. In the study by \cite{andersson_bursts_2023}, volunteers were asked to review the light curves and radio images of ThunderKAT sources and label them as stable, extended, transient/variable, artefacts, or uncertain. We used the transient/variable candidates flagged by citizen scientists as ground truth labels.

For this comparison, we again used the na\"{i}ve thresholds of $\sigma_\textrm{M32} > 0.67$ and $\sigma_\textrm{SE} > 0.67$ for delineating interesting candidates from non-interesting ones. The third column of Table~\ref{tab:sensitivities} summarises this result, where 123 of the 126 sources identified by \cite{andersson_bursts_2023} sit above this threshold. This results in a classifier with a {\emph{true positive rate} or \emph{sensitivity} of 97.6\%. Only three sources were not correctly identified as variable or transient.

The left-hand panel of Figure \ref{fig:sigma-etav-labelled} shows this result graphically, where colour coding indicates whether citizen scientists voted a particular source as a probable transient/variable or not. In this plot, we observe that all the sources marked as variable/transient have values of either $\sigma_\textrm{M32}$ or $\sigma_\textrm{SE}$ (or both) that sit outside the `stable source' region discussed earlier. The dashed lines indicate the 0.67 thresholds used in this classification, and we can observe three sources in the lower left quadrant that were missed by this naive thresholding method, but only marginally so.

These results confirm that our approach is at least consistent with the visual judgement of human classifiers and affirm that locating light curves in the $(\sigma_\textrm{M32},\sigma_\textrm{SE})$ parameter space is a promising method for identifying variable and transient candidates. 

\begin{table}[]    
\caption{Comparison of detection rates of variable/transient sources, treating the 126 candidates identified by citizen scientists in \cite{andersson_bursts_2023} as ground truth (second column from the left). Amplitude hyperparameter estimates $(\sigma_\textrm{M32}, \sigma_\textrm{SE})$ are thresholded at the rule-of-thumb value of $0.67$, that is, sources with either hyperparameter larger than this value are considered scientifically interesting. The $(\eta_\nu, V_\nu)$ metrics are thresholded according to the minimum values used by \cite{rowlinson_identifying_2019}. The approach using amplitude hyperparameters has a much higher sensitivity of detection (97.6\%) compared with the well-established $(\eta_\nu, V_\nu)$ method (56.3\%).}
    \label{tab:sensitivities}
    \centering
    \begin{tabular}{lccc}
    \toprule
    \multirow{2}{2.5cm}{Classification} & \multirow{2}{3.5cm}{\centering \cite{andersson_bursts_2023}} & \multirow{2}{2.5cm}{\centering $\sigma_\textrm{M32} > 0.67$ or $\sigma_\textrm{SE} > 0.67$} & \multirow{2}{1.7cm}{\centering $\eta_\nu > 8.22$ or $V_\nu > 0.179$}\\
    \\
    \midrule
    Transient/variable & $126$ $(100\%)$ & $123$ $(97.6\%) $ & $71$ $(56.3\%)$ \\
    Non-transient/variable & $0$ & $3$ $(2.4\%)$ & $55$ $(43.7\%)$ \\
    \bottomrule
    \end{tabular}
\end{table}

\begin{figure}
\centering
\includegraphics[width=\textwidth]{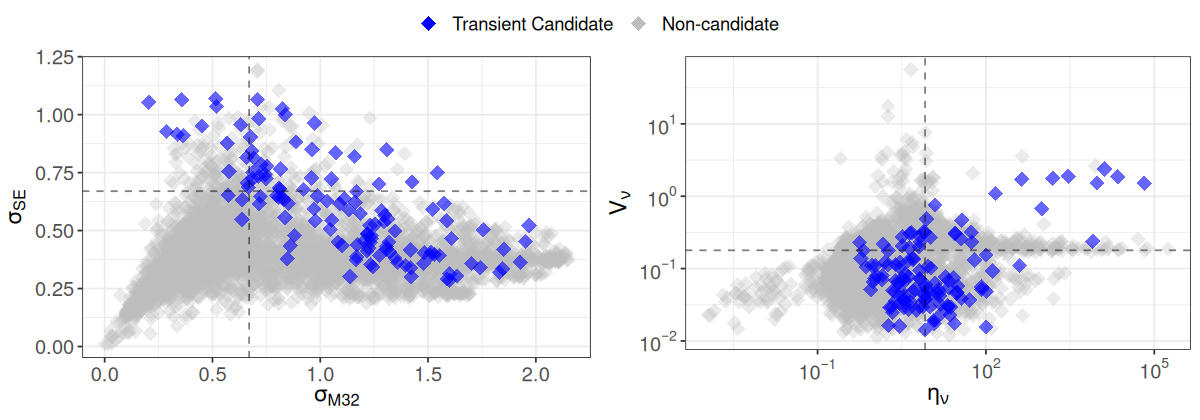}
\caption{\label{fig:sigma-etav-labelled}Left: Scatter plot of the posterior medians of the amplitude hyperparameter of the squared exponential and Matern~3/2 kernels, where each point corresponds to a ThunderKAT light curve. Amplitudes are in standardised units, with dashed lines drawn at $\sigma_. = 0.67$. Right: Scatter plot of the $(\eta_\nu, V_\nu)$ statistics for each ThunderKAT light curve with dashed lines at $8.22$ and $0.179$, respectively.
Blue points are sources voted as probable variable or transient candidates by citizen scientists in \cite{andersson_bursts_2023}. Grey points are sources that were not voted as variable or transient. Searches for transient or variable sources would typically avoid starting in the lower left quadrants of these parameter spaces.}
\end{figure}

\subsubsection{Comparison with $\eta_\nu$ and $V_\nu$}

It is instructive to compare the discriminatory power and classification accuracy of using $(\sigma_\textrm{M32},\sigma_\textrm{SE})$ with the more commonly used variability measures of $\eta_\nu$ and $V_\nu$ (Section~\ref{subsec:eta-V}). We first explore the discriminatory power of these metrics by returning to the same four light curves plotted in Figure~\ref{fig:egs-postpreds}. The right-hand panel of Figure~\ref{fig:sigma-M32-SE-examples} is akin to the left-hand panel except that each light curve is now plotted according to its $(\eta_\nu,V_\nu)$ coordinates. Despite being visually very distinct in appearance, these light curves are all situated closely together in $(\eta_\nu, V_\nu)$-space and not in the upper-right quadrant where searches for variable sources have traditionally focused. The nominally stable light curve labelled \textsf{C} is somewhat set apart from the other three, but in Table~\ref{tab:etanu-comparison}, we see that all four have similar $\eta_\nu$ and $V_\nu$ values. This contrasts with their location in the left-hand panel of Figure~\ref{fig:sigma-M32-SE-examples}, where each source is well separated and points to the  $(\sigma_\textrm{M32},\sigma_\textrm{SE})$ characterisation having stronger discriminatory power.

Figure~\ref{fig:xrbs-scatter} in Appendix~\ref{app:known-xrbs} explores this relationship in the reverse direction by plotting the location in $(\sigma_\textrm{M32},\sigma_\textrm{SE})$-space of sources identified as highly variable in $(\eta_\nu, V_\nu)$-space. We observe that all six examples, each a known variable X-ray binary system, lie far away from the region of stable sources described in Section~\ref{subsec:stable-sources}, consistent with their classification based on their $(\eta_\nu, V_\nu)$ values.

As was done for $\sigma_\textrm{M32}$ and $\sigma_\textrm{SE}$ in Section~\ref{subsubsec:label_comparison}, we revisit the citizen scientist results from \cite{andersson_bursts_2023} and compare them with an approach based on thresholding $\eta_\nu$ and $V_\nu$. For thresholds, we adopt $\eta_\nu > 8.22$ and $V_\nu > 0.179$, as used by \cite{rowlinson_identifying_2019} in their identification of variable sources. The results are shown in the rightmost column of Table~\ref{tab:sensitivities}, where we see correct identification of 71 of the 126 sources reported by \cite{andersson_bursts_2023}. This amounts to a sensitivity of $56.3\%$, which is much lower than the $97.6\%$ sensitivity afforded by classifications based on thresholding $\sigma_\textrm{M32}$ and $\sigma_\textrm{SE}$. 

Naturally, we could boost the sensitivity of any classifier by simply lowering the thresholds; however, this comes at the cost of reducing a classifier's \emph{specificity}. A classifier with poor specificity will return many false positives by flagging spurious candidates that are more likely to be non-variable than variable. The panels in Figure~\ref{fig:sigma-etav-labelled} illustrate this trade-off. Whilst large values of $\eta_\nu$ and $V_\nu$ in the right-hand panel are consistent with a source being classified as transient (coloured points), lowering the thresholds for $\eta_\nu$ and $V_\nu$ to identify all these transients correctly would also include most of the other sources \emph{not} flagged by \cite{andersson_bursts_2023} as likely transients. This is not the case in $(\sigma_\textrm{M32},\sigma_\textrm{SE})$-space, since transient candidates are largely confined to the upper-right quadrant of the parameter space. 

These comparisons show that our GP-based characterisation has a parameter space that better segments variable and transient sources than the traditional ($\eta_\nu$, $V_\nu$)-space.

\subsection{Accounting for field of view effects}
\label{subsec:survey-effects}

The statistical characterisation of an astronomical source based on its observed properties assumes that these are a reliable and uncontaminated proxy of its true physical properties. However, in all surveys, including ThunderKAT, we expect additional effects related to the observing regime rather than the source itself. Consequently, any fitted hyperparameters will also reflect these contaminants. For example, observing across different fields of view, each with different numbers of pointings, over multiple epochs, has manifested in the distinct binning and clustering of sources by colour (field) in the scatter plots in Figure~\ref{fig:hyperparam}. 

Observational cadence is an obvious property that distinguishes fields from one another and is a potential explanation for the binning effects described above. In the absence of true astrophysical periodicity, the periodic kernel term of our GP regression model might be expected to converge onto periodic patterns such as observational cadence. However, the coloured clusters in the posterior medians of $T$ seen in Figure~\ref{fig:hyperparam}\textsf{D} are not aligned with the known observation cadences, and inspection of their posteriors showed very little change from their hyperpriors. Refitting the data without the periodic kernel term also yielded similar posteriors. This suggests that the periodic kernel has detected neither the survey cadence nor any other periodic component. Having ruled out cadence, further investigation revealed that the duration of the light curve is the main confounding factor affecting the length scale hyperparameter estimates ($\ell$). This is clearly shown in the observations around GX 339-4, coloured yellow in Figures~\ref{fig:hyperparam} and~\ref{fig:per-T-ell-T-N}, which is the field with the longest duration light curves and whose hyperparameter estimates are clustered at the largest time scales.

We notice, however, that in Figure~\ref{fig:sigma-M32-SE} there are comparatively fewer binning artefacts in the scatter plot of the amplitude hyperparameters, $\sigma_\textrm{M32}$ and $\sigma_\textrm{SE}$. This implies that these two measures are less impacted by confounding by the duration of light curves and are more reliable for characterising sources. However, should data uncertainties be too large, observing campaigns too short, or cadence too sparse relative to the timescale of interest, such light curves will likely be situated in the non-varying region described in Section~\ref{subsec:stable-sources}. 

\subsection{Transient Candidates}
\label{sec:transient_candidates}

Following our claim that the $(\sigma_\textrm{M32},\sigma_\textrm{SE})$ hyperparameter space is more favourable than the $(\eta_\nu$,$V_\nu)$ parameter space for identifying variables and transients, we conducted an elementary search for candidate sources in the ThunderKAT dataset using our GP regression approach. Searching the hyperparameter space near known transient sources is a natural place to start, and the most obvious targets are the known X-ray binary (XRB) systems upon which ThunderKAT fields of view are centred. The light curves and characteristics of the six XRBs used to `seed' our search are described in Appendix~\ref{app:known-xrbs}. Rather than a rigorous search for transient sources, we are demonstrating how our proposed metrics might be incorporated as an initial single-pass screening step in a more involved search protocol.

A basic search algorithm is as follows: 
\begin{enumerate}
    \item For a known seed XRB source in the ThunderKAT dataset.
    \item Select its 30 nearest neighbours in the $(\sigma_\textrm{M32},\sigma_\textrm{SE})$ hyperparameter space by Euclidean distance, excluding sources from the same field as the original known XRB. The reason for excluding sources from the same field is to ensure that any statistical similarity between sources is not simply due to the similarity in their observing conditions. 
    \item For each potential candidate, visually inspect its radio image with the highest signal-to-noise ratio. If the source is resolved or otherwise ambiguous, remove it from consideration.
    \item For any remaining candidates, visually inspect their light curves and only retain those showing transient behaviour.
\end{enumerate}

We expect any candidate sources we identified to have light curves that are statistically, if not astronomically, similar to the target sources. However, as a non-parametric approach, a GP regression model that is not tuned to a particular type of source, say XRBs, may quantify the observed variability patterns of other source types, such as flaring stars, and situate them in the hyperparameter space accordingly. Further work is needed to explore the distribution of different source types within the $(\sigma_\textrm{M32},\sigma_\textrm{SE})$ hyperparameter space, as well as the impact of using different types of sources to seed searches.

Thirty sources were identified from each of the six known XRBs for an initial list of 180 candidates. After removing sources already identified by citizen scientists as transients in \cite{andersson_bursts_2023}, visual inspection of radio images and light curves was conducted to remove extended sources and observations with spurious data points such as negative flux densities. The supplementary materials contain a data file with details of the remaining 67 unique candidates. For illustrative purposes, ten candidates' light curves and fitted models are plotted in Figure~\ref{fig:candidates-postpred} and their variability metrics summarised in Table~\ref{tab:transient-candidates}. This sample of ten was deliberately chosen to showcase light curves from a broad selection of observing fields.

\begin{table}

\caption{Location and variability metrics for the ten candidate sources shown in Figure~\ref{fig:candidates-postpred}. Entries are sorted in ascending size of $\sigma_\textrm{M32}$. None of the $(\eta_\nu,V_\nu)$ values are remarkable.}
\centering
\label{tab:transient-candidates}
\begin{tabular}{clllrccccc}
\toprule
\multirow{2}{*}{Figure~\ref{fig:candidates-postpred}} & \multirow{2}{*}{\centering Seed XRB} & \multicolumn{8}{c}{Candidate Source} \\
& & Field & RA ($^\circ$) & Dec. ($^\circ$) & Mean Flux (mJy) & $\eta_\nu$ & $V_\nu$ & $\sigma_\textrm{M32}$ & $\sigma_\textrm{SE}$\\ 
\midrule
a & H~1743$-$322 & \textsf{4U1543} & 236.9659 & -47.6767 & 0.410 & 3.48 & 0.17 & 0.59 & 0.90\\
b & MAXI~J1820$+$070 & \textsf{GX339} & 256.6843 & -48.9579 & 0.804 & 1.54 & 0.09 & 0.67 & 0.66\\
c & H~1743$-$322 & \textsf{MAXIJ1803} & 271.1269 & -30.4385 & 0.649 & 2.58 & 0.09 & 0.68 & 0.78\\
d & H~1743$-$322 & \textsf{J1858} & 284.3228 & -8.3813 & 0.728 & 1.08 & 0.05 & 0.74 & 0.85\\
e & GX~339$-$4 & \textsf{4U1543} & 237.2070 & -48.3865 & 0.835 & 4.06 & 0.08 & 1.09 & 0.55\\
f & GX~339$-$4 & \textsf{MAXIJ1803} & 271.2300 & -29.6591 & 2.522 & 4.68 & 0.03 & 1.12 & 0.58\\
g & GX~339$-$4 & \textsf{J1858} & 284.9503 & -8.6815 & 0.928 & 1.46 & 0.05 & 1.17 & 0.53\\
h & MAXI~J1803$-$298 & \textsf{EXO1846} & 282.3311 & -2.5805 & 0.291 & 2.29 & 0.47 & 1.33 & 0.42\\
i & MAXI~J1820$+$070 & \textsf{MAXIJ1348} & 207.6484 & -62.6685 & 1.306 & 2.41 & 0.09 & 1.62 & 0.37\\
j & MAXI~J1348$-$630 & \textsf{GRS1915} & 289.2072 & 11.0012 & 0.169 & 7.77 & 0.44 & 1.85 & 0.39\\
\bottomrule
\end{tabular}
\end{table}

\begin{figure}
\centering
\includegraphics[width=0.97\linewidth]{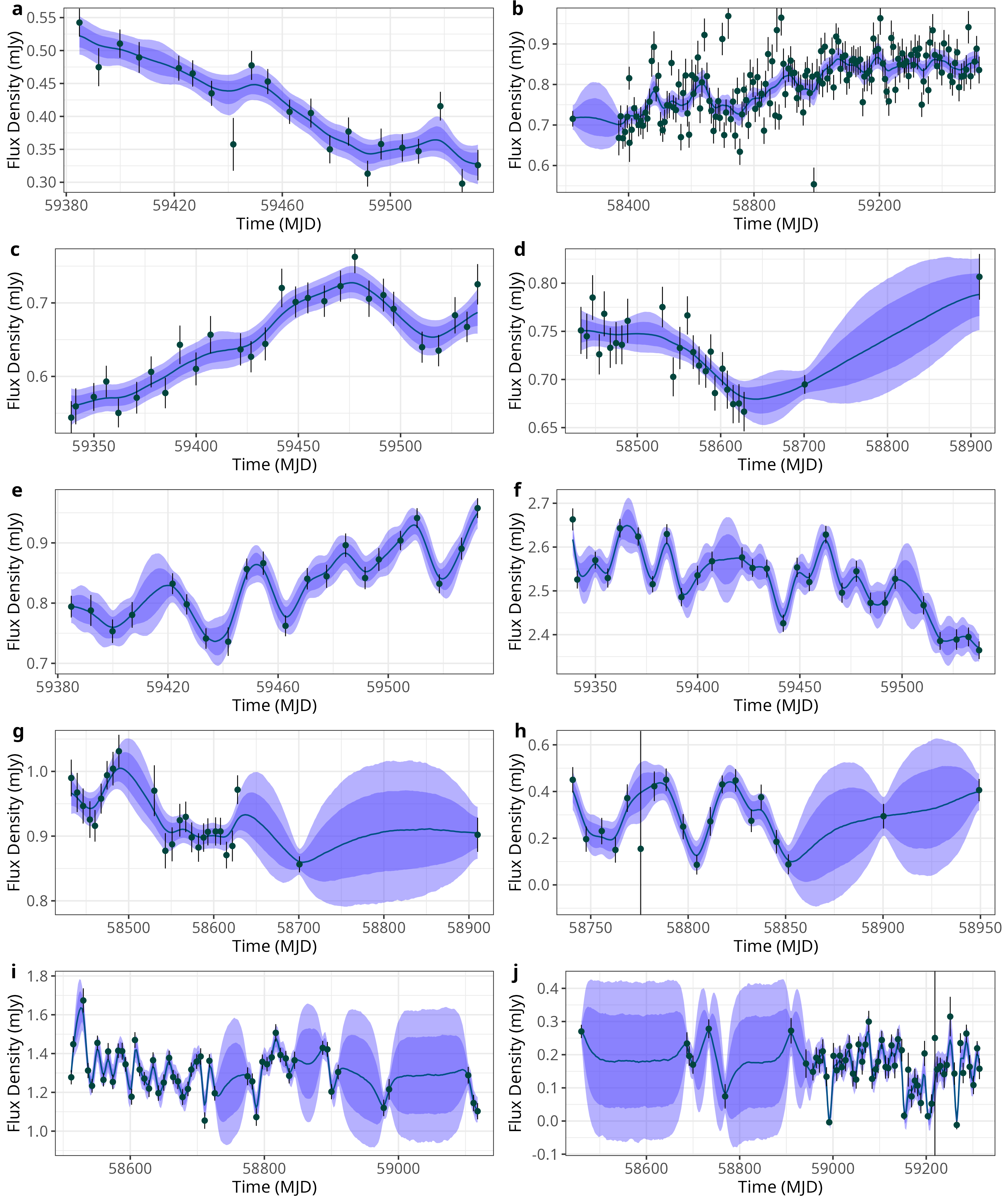}
\caption{\label{fig:candidates-postpred}Fitted light curve models for ten transient candidates identified using a basic search algorithm. None of these sources were voted as candidate transient or variable sources by citizen scientists in \cite{andersson_bursts_2023}. Location and variability statistics for each source are shown in Table~\ref{tab:transient-candidates}. The `seed' light curves to which these candidates were adjacent are shown in Figure~\ref{fig:xrb-postpreds}.}
\end{figure}

The ten candidate light curves shown in Figure~\ref{fig:candidates-postpred} are ordered by increasing $\sigma_\textrm{M32}$, the hyperparameter tuned for detecting short timescale variability. This is reflected by light curves in the top panels exhibiting long timescale trends and fewer undulations, compared with the lower panels showing more jagged appearances. Also, while the seed light curves shown in Figure~\ref{fig:xrb-postpreds} appear burst or flare-like, these ten candidate sources behave more like persistent variables. Despite these differences in shape, all of these light curves have median $\sigma_\textrm{SE}$ and $\sigma_\textrm{M32}$ estimates greater than 0.67, identifying them as varying sources. The fact that this model can capture different types of variability reflects the flexibility of GP models and points to our model's generalisability to a diverse range of source types.

Figure~\ref{fig:candidates-scatter} compares where these ten candidate sources sit in two respective parameter spaces. Once again, we observe that the separation of these candidates from both each other and from nominally non-varying sources is more pronounced in the $(\sigma_\textrm{M32},\sigma_\textrm{SE})$-space compared with the $(\eta_\nu$,$V_\nu)$-space. Also, the grouping of light curves within the amplitude hyperparameter space seems more interpretable since sources whose light curves are visually similar, such as sources \textsf{a}--\textsf{d}, are situated away from source \textsf{j} whose behaviour is clearly distinct.

\begin{figure}
\centering
\includegraphics[width=1\linewidth]{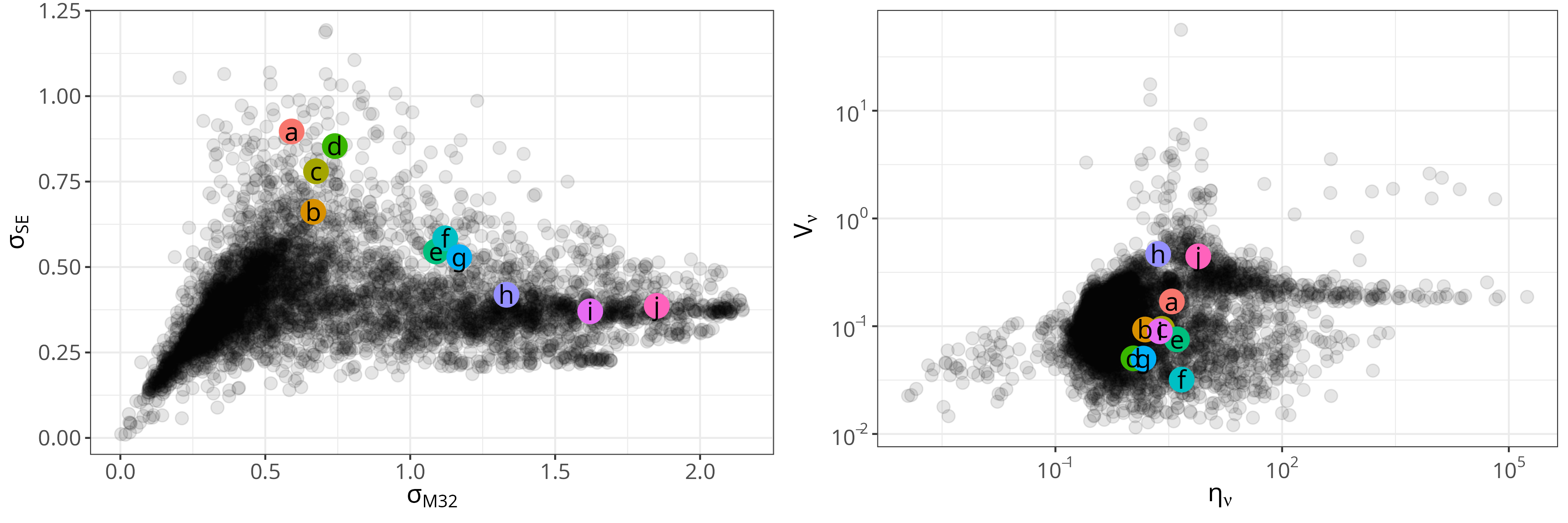}    
\caption{\label{fig:candidates-scatter}Left: The $(\sigma_\textrm{M32},\sigma_\textrm{SE})$ locations of ten candidates identified by searching the statistical neighbourhood of the six known XRB sources labelled in Figure~\ref{fig:xrbs-scatter}. Right: The location in $(\eta_\nu$,$V_\nu)$ of the same sources shown in the left panel. Notice that the locations of these sources in $(\eta_\nu$,$V_\nu)$ are distant from the known XRB sources as shown in the corresponding panel of Figure~\ref{fig:xrbs-scatter}. Coordinates and variability metrics of each source are summarised in Table~\ref{tab:transient-candidates}.}
\end{figure}

\subsection{Proposed extensions}

Several simplifications that we have made require further elaboration. The first is the stationarity of light curves. This strong assumption overlooks situations in which light curves are composites of multiple underlying astrophysical processes, each contributing different components of variability at different phases of activity. This shortcoming could be mitigated by relaxing our modelling choice of using a zero mean function. However, using a non-zero, parametric mean function based on prior knowledge may bias the GP regression model toward particular astrophysical phenomena. One approach is to use a non-parametric function, perhaps even another GP, as the mean function of a more complex hierarchical model.

The second is that we do not postulate a probability model for the observational uncertainty but instead directly use the provided standard errors. Considerable work would be needed to specify the distribution of the observational uncertainties in radio interferometric data. This may entail incorporating work on non-linear variability found in X-ray light curves \citep{uttley_non-linear_2005}, which may also moderate the effect of outliers such as those seen in Figures~\ref{fig:postpreds_psds}{\textsf{G}} and \ref{fig:egs-postpreds}\textsf{D}, whose uncharacteristically small standard errors heavily distort the fitted model.

Thirdly, we have used point values (medians) to summarise the posterior distribution of each hyperparameter. A more complete analysis would use posterior distributions.

%% file: conclusion.tex
\section{Conclusion} 
\label{sec:conclusion}

In this work, we have presented an approach using Gaussian process regression to characterise the variability of sources based on their light curves. Gaussian process models such as ours implicitly handle non-uniform data coverage and cadence, which are properties that are commonplace in time-domain astronomical surveys. We showed that our approach offers greater discriminatory power than the commonly used parameterisation of $(\eta_\nu, V_\nu)$. Furthermore, it differs from existing applications of Gaussian processes in the astronomical literature because it adopts a multi-term kernel function. We specify a composite covariance kernel of squared exponential, Matern~3/2, and periodic terms. Then, we use the posterior estimates of their hyperparameters, not the mean posterior predictive curve, to distinguish between variable and non-variable sources. Nevertheless, the mean curves and their PSDs are useful diagnostics for model assessment. 

Our key finding is that the bivariate distribution of the amplitude hyperparameters of the squared exponential and Matern~3/2 kernels, namely $\sigma_\textrm{SE}$ and $\sigma_\textrm{M32}$, defines a parameter space where statistically different light curves populate distinctly different regions. Qualitatively, this means that sources exhibiting high variability on long- or short-time scales (or both) are clustered away from sources whose light curves are nominally non-varying or stable. This is very useful in screening for transient or variable sources. The astronomer interested in identifying variable or transient source candidates should examine light curves whose $\sigma_\textrm{SE}$ or $\sigma_\textrm{M32}$ are in the upper quantiles of their survey, since these should prove the most promising. We demonstrated this by identifying sources in ThunderKAT data that were not previously flagged in the literature. We claim this is a generalisable approach that can be used to discover transient and variable candidates, agnostic of source type. Having tested this method on one survey with reasonable results, applying this approach to other large surveys is the subject of upcoming work.

Computational notebooks in \textsc{R} and \textsc{Python} describing our method are available\footnote{hosted at \url{https://github.com/shihchingfu/fu2025_transients_gp} and preserved on Zenodo~\citep{fu_computational_2024}} to help facilitate its translation and application to surveys other than ThunderKAT.

%% file: appendix.tex
\section{Justification of Hyperpriors}
\label{app:hyperpriors}

Given the anticipated diversity of objects observed in a large survey such as ThunderKAT, it is unlikely that the same kernel hyperparameter values would be suitable across all sources. Yet, some physical constraints should be consistent regardless of the underlying astrophysical phenomena observed. Below, we elaborate on how we express such constraints in terms of kernel hyperprior distributions and justify the choice of these hyperpriors (Eqs.~\ref{eq:sigma_prior}--\ref{eq:T_prior}).

\paragraph{Length scale} We assigned identical hyperpriors to each kernel's length scale hyperparameter ($\ell$). As suggested by the \textsc{Stan} User's Guide \citep[][p.~153]{stan_development_team_stan_2023}, we used an Inverse Gamma (IG) hyperprior since it down-weights very small length scales -- scales that are more comparable to noise than signal -- whilst still permitting medium to large length scales. The shape parameter of the IG distribution is fixed at $\alpha = 3$, and the rate parameter $\beta$ is set to one-half the range of the light curve duration (Eq.~\ref{eq:ell_prior}).

Recalling that the expected value of an IG distributed random variable is $\beta / (\alpha - 1)$, the above initialises the prior mean of these length scale hyperpriors to approximately one-quarter of the total period of the light curve. This is a good rule of thumb since we want to observe at least four cycles of repeating patterns before judging any pattern as real, not noise. Additionally, length scales were given a lower bound of the shortest separation in time between data points, $\Delta t$ (Eq.~\ref{eq:ell_min}). This is because it is unreasonable to infer autocorrelation structures shorter than the finest resolution in the data. Thus, this lower bound prevented length scales from approaching zero too closely.

In addition to a general lower bound for all length scale hyperparameters, a further inequality was placed on $\ell_\textrm{M32}$ such that it must be smaller than $\ell_\textrm{SE}$ (Eq.~\ref{eq:ell-SE-M32-constraint}). This mitigates identifiability issues arising from the similar behaviour of these kernels when applied at similar scales. Depending on the light curve, either kernel is equally capable of describing the same autocovariance structures. By restricting the squared exponential kernel to take on larger length scales, the kernel is impelled to lock onto longer and smoother trends for which it is better suited. Thus, the Matern~3/2 kernel tends to capture more jagged patterns found at shorter length scales.

\paragraph{Amplitude} For amplitude hyperparameters ($\sigma$), standard half-normal hyperpriors were applied to the standardised flux densities (Eq.~\ref{eq:sigma_prior}). Indeed, flux densities were standardised to facilitate applying identical amplitude hyperpriors to all light curves, which otherwise have different flux density ranges. A useful rule-of-thumb for assessing whether a posterior hyperparameter estimate is significant is comparing it with a nominal threshold derived from its hyperprior distribution. The median of the standard half-normal distribution is $\approx 0.67$, which is therefore a useful threshold for gauging the influence of this hyperprior on the resultant posterior -- values above 0.67 warrant further investigation.

\paragraph{Period} A uniform hyperprior was chosen for the period ($T$) hyperparameter since we do not have any general information about the presence or absence of periodicity in ThunderKAT light curves. A uniform hyperprior will avoid biasing the detection of periodicities towards any particular period, and any period in the range is plausible. As with the amplitude hyperpriors, the sparsity of observations was used to inform the bounds of the uniform hyperprior (Eq. \ref{eq:T_prior}).

Similar to the length scale hyperpriors, we wanted to observe at least three or four full periodic cycles in our data before concluding there was any evidence of periodicity. This heuristic sets an upper bound on the uniform hyperprior to be one-quarter of the total time span of the light curve. For a lower bound, we took the shortest time gap between observations ($\Delta t$), or conversely, the highest instantaneous sampling rate, and multiplied by two to get the Nyquist frequency. This prevented the model from fitting periods smaller than twice the smallest time gap, i.e., never finer than the best data resolution.

\section{Marginal Gaussian Process Model}
\label{app:marginalgp}

In our model, the uncertainties in the flux density measurements are assumed to be Gaussian distributed. This means we can `integrate out' the latent function $f$ and re-express Eqs. \ref{eq:flux-distn} and \ref{eq:flux-gp} more simply as a \textit{marginal} Gaussian process model:
\begin{equation}
\label{eq:marginalgp}
S \sim \mathcal{GP}\left(\boldsymbol{0}, \boldsymbol{K}_{N \times N} + \mathrm{diag}(\boldsymbol{\hat{e}}^2)\right)
\end{equation}

\noindent where $\mathrm{diag}(\boldsymbol{\hat{e}}^2)$ is an $N \times N$ matrix whose diagonal values are the error variances of the observed data, $\boldsymbol{\hat{e}}^2$, and the off-diagonal values are zero. A marginal model is an acceptable simplification because the main object of this study is the characterisation of light curves rather than explicitly inferring the `true' flux density function $f$. This formulation has the added benefit of having an analytical expression for the posterior predictive distribution \cite[][p.~16]{rasmussen_gaussian_2006}: 
\begin{equation}
\label{eq:postpred}
\boldsymbol{s^*} \mid \boldsymbol{t^*}, \boldsymbol{s}, \boldsymbol{t} \sim \mathcal{N}\left(\boldsymbol{K_*^\top} \boldsymbol{\Sigma}^{-1} \boldsymbol{y}, \boldsymbol{\Omega} - \boldsymbol{K_*^\top} \boldsymbol{\Sigma}^{-1} \boldsymbol{K_*}\right)
\end{equation}

\noindent where $\boldsymbol{s}$ are the flux densities observed at times $\boldsymbol{t}$, and $\boldsymbol{s^*}$ are the predicted flux densities at the target times $\boldsymbol{t^*}$. $\boldsymbol{\Sigma}$ is the covariance matrix as defined in Eq.~\ref{eq:kappa}, $\boldsymbol{K_*}$ is the covariance matrix between times $\boldsymbol{t}$ and $\boldsymbol{t^*}$, and $\boldsymbol{\Omega}$ specifies the covariances between the target times, $\boldsymbol{t^*}$. With this analytic expression for the posterior predictive distribution, drawing samples is much more efficient and straightforward computationally.

\section{Fitted Models of Known XRBs}
\label{app:known-xrbs}

\begin{figure}
\includegraphics[width=1\linewidth]{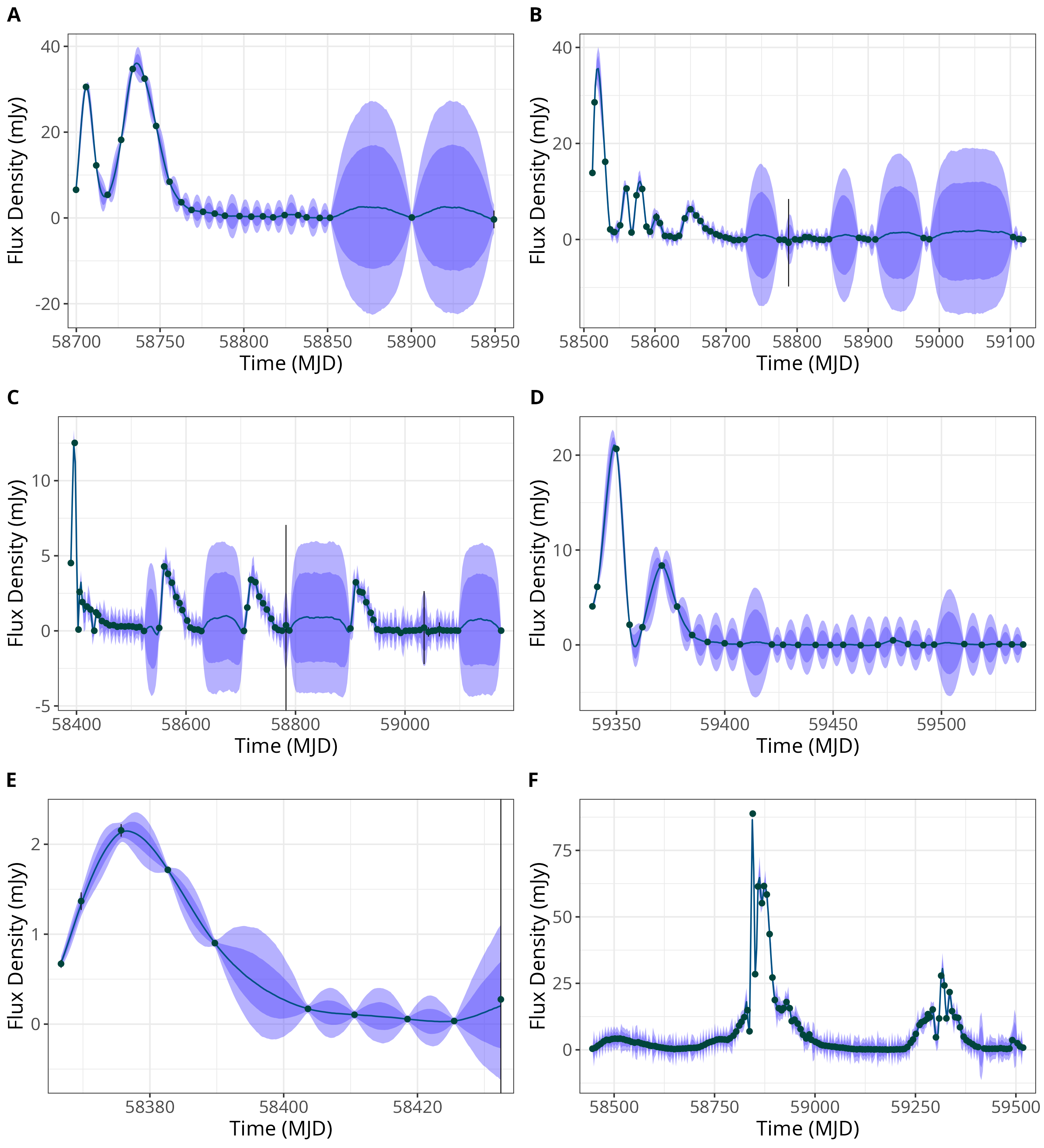}%
\caption{Plots of posterior predictive curves from GP models of light curves of six known X-ray binaries: \textsf{EXO1846} (\textsf{A}), \textsf{MAXIJ1348} (\textsf{B}), \textsf{MAXIJ1820} (\textsf{C}), \textsf{MAXIJ1803} (\textsf{D}), \textsf{H1743} (\textsf{E}), \textsf{GX339} (\textsf{F}). Coordinates of each source and posterior statistics of each light curve are found in Table~\ref{tab:xrb_post_medians}. Points are the original observations, with black bars indicating standard errors. Lines trace the median of the posterior predictive samples, with shaded regions indicating 68\% and 90\% quantile intervals, respectively. }
\label{fig:xrb-postpreds} 
\end{figure}

This section contains more examples of GP regression model fits to light curves. In particular, the six examples shown in Figure~\ref{fig:xrb-postpreds} are light curves of sources known to be X-ray binary (XRB) systems, namely: \textsf{EXO1846}, \textsf{MAXIJ1348}, \textsf{MAXIJ1820}, \textsf{MAXIJ1803}, \textsf{H1743}, and \textsf{GX339}. Indeed, these are the sources at the centre of the various fields of view observed in the ThunderKAT survey. Each source exhibits transient behaviour, namely large spikes in flux density before resuming some previous level. This makes them suitable cases for testing whether our model can identify transient-like sources.

\begin{figure}
\centering
\includegraphics[width=1\textwidth]{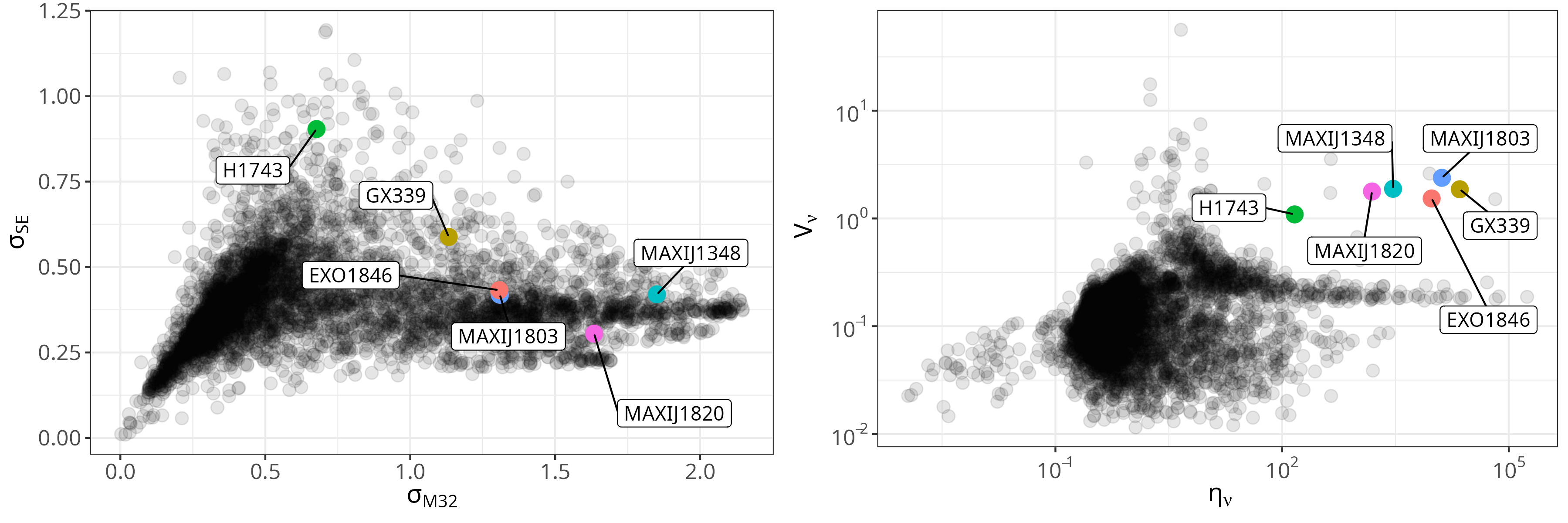}
\caption{\label{fig:xrbs-scatter}Left: Scatter plot of the posterior medians of the amplitude hyperparameter of the squared exponential and Matern~3/2 kernels. Amplitudes are in standardised units. Right: Scatter plot of the $(\eta_\nu$,$V_\nu)$ values for the same light curves. In both panels, six known XRBs are marked by coloured points. }
\end{figure}

Table~\ref{tab:xrb_post_medians} summarises the variability statistics and estimated GP regression hyperparameters for each of these XRB light curves. In every case, the posterior medians of $\sigma_\textrm{M32}$ are in regions of hyperparameter space that indicate non-negligible variability. This is clearly shown in the scatter plot of Figure~\ref{fig:xrbs-scatter}, where these sources are marked by coloured points.

\begin{table}
\caption{\label{tab:xrb_post_medians}Fitted hyperparameter values for the example light curves shown in Figure~\ref{fig:xrb-postpreds}; each of which are known X-ray binary systems. Medians of the hyperparameter posteriors are shown beside the frequently reported variability statistics of $\eta_\nu$ and $V_\nu$. Boldface indicates median amplitudes greater than 0.67.}
\centering
\begin{tabular}{clccccccccccc}
\toprule
\multirow{2}{1.5cm}{\centering Figure~\ref{fig:xrb-postpreds}} & \multirow{2}{*}{XRB} & \multirow{2}{*}{RA ($^\circ$)} & \multirow{2}{*}{Dec. ($^\circ$)} & \multirow{2}{*}{$\eta_\nu$}& \multirow{2}{*}{$V_\nu$} & \multicolumn{7}{c}{Posterior Median} \\
& & & & & & $\sigma_\textrm{M32}$ & $\ell_\textrm{M32}$ & $\sigma_\textrm{SE}$ & $\ell_\textrm{SE}$ & $\sigma_\textrm{P}$ & $\ell_\textrm{P}$ & $T$ \\
\midrule
\textsf{A} & \textsf{EXO1846} & 282.3212 & -3.0655 & 9528.87 & 1.53 & \textbf{1.31} & 13.89 & 0.43 & 48.39 & 0.48 & 37.44 & 86.12\\
\textsf{B} & \textsf{MAXIJ1348} & 207.0532 & -63.2747 & 2936.57 & 1.88 & \textbf{1.85} & 16.26 & 0.42 & 119.09 & 0.48 & 90.54 & 209.01\\
\textsf{C} & \textsf{MAXIJ1820} & 275.0914 & 7.1853 & 1546.85 & 1.78 & \textbf{1.63} & 9.45 & 0.31 & 160.82 & 0.41 & 145.90 & 199.04\\
\textsf{D} & \textsf{MAXIJ1803} & 270.7617 & -29.8301 & 13026.88 & 2.38 & \textbf{1.31} & 9.07 & 0.42 & 40.96 & 0.47 & 29.52 & 68.22\\
\textsf{E} & \textsf{H1743} & 266.5652 & -32.2340 & 146.35 & 1.09 & \textbf{0.68} & 9.36 & \textbf{0.90} & 13.98 & 0.49 & 10.06 & 23.38\\
\textsf{F} & \textsf{GX339} & 255.7055 & -48.7897 & 22427.60 & 1.86 & \textbf{1.13} & 10.75 & 0.59 & 127.85 & 0.40 & 159.48 & 368.66\\
\bottomrule
\end{tabular}
\end{table}

\section{Principal Components Analysis of Hyperparameter Posterior Means}
\label{app:PCA}

Principal component analysis (PCA) of the posterior means of the fitted hyperparameters for every light curve reveals that two principal components (PCs) are sufficient to explain over 77\% of the variability in the data. A third component increases this to over 94\% (see the second column of Table~\ref{tab:pca-loadings}). The loadings for the length scale $\ell$ and period $T$ hyperparameters are negative, which shows that these hyperparameters are negatively correlated with the first PC. Conversely, the amplitude $\sigma$ hyperparameters are positively correlated with the second principal component. This suggests that the first PC, which increases as the length scale and period decrease, captures short time-scale variability in the light curves. Such variability is likely to be noise rather than phenomenological, with the remaining variability explained by the amplitude hyperparameters of the second PC.

\begin{figure}
    \centering
    \includegraphics[width=0.7\textwidth]{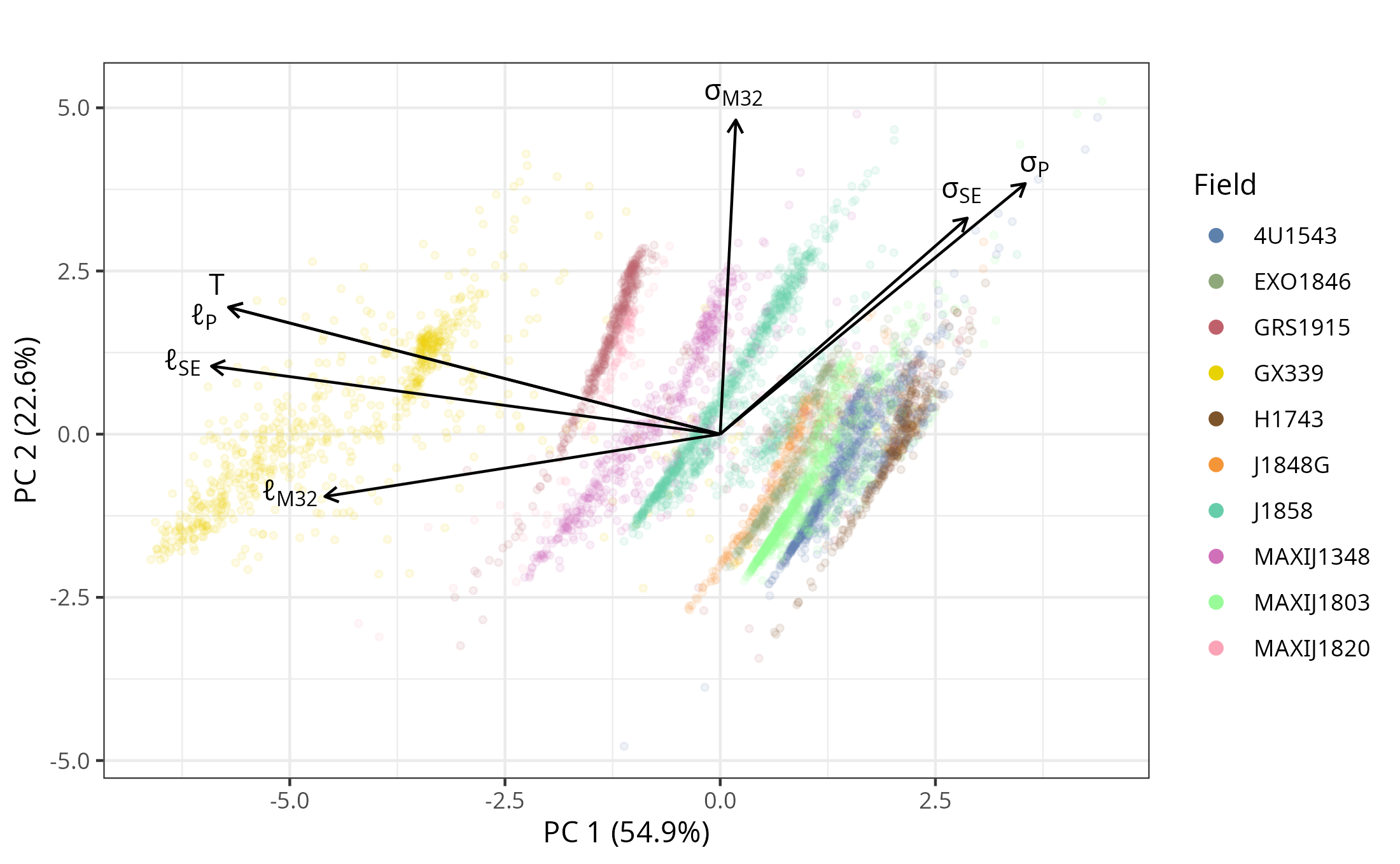}%
    \caption{Biplot of the first two principal components (PCs) of the posterior means of the hyperparameters estimated for each light curve. Hyperparameter values were standardised before applying principal components analysis. Colours indicate in which field the light curve was observed, and the arrows indicate the loadings of the original hyperparameters projected onto these first two principal components. The $\ell_\mathrm{P}$ and $T$ hyperparameters are perfectly correlated, so their arrows overlap.}
    \label{fig:pca-biplot} 
\end{figure}

The biplot in Figure~\ref{fig:pca-biplot} shows the position of each light curve as projected into the 2D plane of the first two principal components. The arrows are scaled representations of the loadings in Table~\ref{tab:pca-loadings}, and the colours indicate the field of observation for each light curve. We observe that the progression of colours (fields) from left to right corresponds strongly to the direction of the first PC, which reflects the confounding of these hyperparameters by field-specific properties, such as the typical duration of those observations. The spread of points within each field towards the top-right is consistent with the loadings of the amplitude hyperparameters of the Matern 3/2 and squared exponential kernels, noting that we will ignore the amplitude hyperparameter of the periodic kernel since it is strongly correlated with the squared exponential.

From these results, we are confident that examining the two hyperparameters of $\sigma_\mathrm{M32}$ and $\sigma_\mathrm{SE}$ is sufficient to characterise light curves fitted by our GP regression model.

\begin{table}
    \caption{Proportion of variance explained by each principal component (PC) and the loadings of each hyperparameter against the first three PCs. Non-negligible values are in boldface. Notice that the first PC is dominated by length scale hyperparameters and the second PC by amplitude hyperparameters.}
    \label{tab:pca-loadings}
    \begin{tabular}{ccccccccc}
        \toprule        
        \multirow{2}{1.5cm}{\centering Principal Component} & \multirow{2}{1.5cm}{\centering Proportion of variance} & \multicolumn{7}{c}{Loadings}\\
        & & $\sigma_\mathrm{SE}$ & $\sigma_\mathrm{M32}$ & $\sigma_\mathrm{P}$ & $\ell_\mathrm{SE}$ & $\ell_\mathrm{M32}$ & $\ell_\mathrm{P}$ & $T$ \\
        \midrule
        1 & 54.9\% & 0.241 & 0 & 0.297 & \textbf{-0.496} & \textbf{-0.385} & \textbf{-0.479} & \textbf{-0.479} \\
        2 & 22.6\% & \textbf{0.433} & \textbf{0.630} & \textbf{0.502} & 0.136 & -0.125 & 0.254 & 0.254 \\
        3 & 16.8\% & 0.559 & -0.528 & 0.325 & 0.127 & 0.532 & 0 & 0\\
        \bottomrule
    \end{tabular}

\end{table}

